\documentclass[preprint,11pt]{JHEP3} 

\JHEPspecialurl{http://jhep.sissa.it/JOURNAL/JHEP3.tar.gz}

\usepackage{epsfig,multicol,amsmath}

\usepackage{rotating}

\usepackage{cite}

\def\beq{\begin{equation}}
\def\eeq{\end{equation}}
\def\bea{\begin{eqnarray}}
\def\eea{\end{eqnarray}}

\def\eq#1{{Eq.~(\ref{#1})}}
\def\fig#1{{Fig.~\ref{#1}}}

\newcommand{\Lb}{\left(}
\newcommand{\Rb}{\right)}
\parskip=\itemsep               

\newcommand{\nn}{\nonumber}

\newcommand{\h}{\frac{1}{2}}

\newcommand{\Ga}{\Gamma}

\renewcommand{\thefigure}{{\protect\bf\arabic{figure}}}

\def\pom{{I\!\!P}}

\title{Long range rapidity correlations {\bf in} soft interaction at high 
energies. }
\author{\Large  E. Gotsman$^{a}$\thanks{Email:
gotsman@post.tau.ac.il.}\,, E. Levin$^{a,b}$\thanks{Email:
leving@post.tau.ac.il}\,\,\,\,and\,\,U. Maor$^{a}$
\thanks{Email: maor@post.tau.ac.il.}\, 
\\
a)\,\,Department of Particle Physics, School of Physics and Astronomy,
Raymond and Beverly Sackler
 Faculty of Exact Science, Tel Aviv University, Tel Aviv, 69978, Israel\\ 
b)\,\,Departamento de F\'\i sica, Universidad T\'ecnica Federico Santa 
Mar\'\i a, Avda. Espa\~na 1680\\ and Centro 
Cientifico-Tecnol$\acute{o}$gico de Valparaiso,Casilla 110-V, 
Valparaiso, Chile
\\}

\abstract  { In this paper we  take the next step (following the 
successful description of  inclusive hadron production) in   describing 
 the structure of the 
bias events without the aid of  Monte Carlo codes.
 Two new results are presented :(i)  a method for calculating
  the two particle correlation functions in the BFKL Pomeron calculus
 in zero transverse dimension;
 and (ii)  an estimation of the   values of these correlations in a model 
of  soft interactions.  
Comparison with the multiplicity data at the LHC is given.
}

\dedicated{PACS: 13.85.-t, 13.85.Hd, 11.55.-m, 11.55.Bq}

\preprint{TAUP -2973/13\\
{\tt }\\
\today}

\begin{document}
\section{Introduction}
\par
The goal of this paper is twofold: to  consider the two hadron long range 
rapidity correlations in 
the BFKL Pomeron Calculus in zero transverse
dimensions; and to calculate these correlations in a model of soft 
interactions at high energy. The BFKL Pomeron Calculus in zero transverse
dimension describes the interaction of the Pomerons through 
the triple Pomeron vertex ($ G_{3 \pom}$) with a 
 Pomeron intercept $\Delta_\pom \equiv \Delta > 0$ 
and a Pomeron slope $\alpha'_\pom =0$.  
The theory that includes all these ingredients can be formulated in a
functional integral form \cite{BRN}:
\begin{equation} \label{FI}
Z[\Phi, \Phi^+]\,\,=\,\,\int \,\,D \Phi\,D\Phi^+
\,e^S \,\,\,\,\,\,\,\,\, \mbox{with}\,\,\,\,\,\,\,S \,=\,S_0
\,+\,S_I\,+\,S_E\,,
\end{equation}
where, $S_0$ describes free Pomerons, $S_I$ corresponds to their mutual 
interaction
and $S_E$ relates to the interaction with the external sources (target and
projectile).  Since  $ \alpha^{\prime}_{\pom}\,=\,0$, $S_0$ has the form
\begin{equation} \label{S0}
S_0\,=\,\int d Y \Phi^+(Y)\,\left\{ -\,
\frac{d }{d Y} \,\,
+\,\,\Delta\,\right\} \Phi(Y).
\end{equation}
$S_I$ includes only triple Pomeron interactions and has the form
\beq \label{SI} S_I\,=\,G_{3\pom} \int d Y\,\left\{
\Phi(Y)\,
\Phi^+(Y)\,\Phi^+(Y)\,\,+\,\,h.c. \right\}. \eeq 
For $S_E$
we have local interactions both in rapidity and in impact parameter space,
\begin{equation} \label{SE}
S_E\,=\,-\,\int dY \sum_{i=1}^2\,\left\{
\Phi(Y)\,g_i(b)\,\,+\,\,\Phi^+(Y)\, g_i(b)
\right\},
\end{equation}
where, $g_i(b)$ stands for the interaction vertex with the hadrons 
at fixed $b$.

At the moment this theory has two facets. 
First, it is a toy-model  describing the 
interaction of the BFKL Pomerons in QCD. 
Many problems can be solved analytically in this simple model leading to 
a set of possible scenarios for the 
solution in BFKL Pomeron calculus\cite{BRN,QCDHE,BART,MUCD,BK,KL,JIMWLK}. 
Our first goal is to find an analytical
solution for the correlation function in rapidity defined as
\beq \label{I1}
R\Lb y_1, y_2\Rb\,\,\,=\,\,\,\frac{\frac{1}{\sigma_{in}}\,
\frac{d^2 \sigma}{d y_1\,d y_2}}{\frac{1}{\sigma_{in}}\,
\frac{d \sigma}{d y_1}\,
\frac{1}{\sigma_{in}}\,\frac{d \sigma}{d y_2}}\,,
\eeq
where, $\sigma_{in}$, $d^2 \sigma/d y_1\,d y_2$ and $d \sigma/d y$ 
are inelastic, double and single inclusive cross sections.
We consider this problem as the most natural starting point to search 
for a solution 
for $R\Lb y_1,y_2\Rb$, in 
a more general and more difficult approach based on high density QCD.

On the other hand, recent experience in building models for high 
energy scattering 
\cite{GLMLAST,GLM1,GLM2,KAP,KMR,OST} shows that a Pomeron with 
$\alpha'_\pom \,=\,0$ can 
describe  the experimental data including that at the LHC. It also appears 
in N=4 
SYM \cite{BST,HIM,COCO,BEPI,LMKS}
with a large coupling, which at the moment,  is the only theory that 
allows one to treat the strong interaction on a theoretical basis.
Therefore, our second goal is to evaluate the correlation 
function $R\Lb y_1,y_2\Rb $ in our model 
for soft high energy interactions (see \cite{GLMLAST,GLM1,GLM2}).
\section{Correlation function in the BFKL Pomeron 
Calculus in zero transverse dimensions}
\subsection{General  approach}
\begin{figure}
\begin{center}
\includegraphics[width=0.7\textwidth]{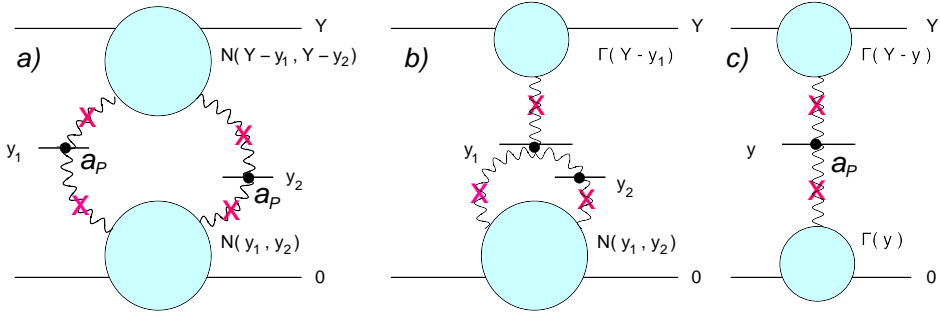}
\end{center}
\caption{The Mueller diagram \cite{MUDI} for double 
(\protect\fig{CorGen}-a and b) and single (\protect\fig{CorGen}-c) 
inclusive cross section. The wavy lines denote Pomerons. 
The cross on a wavy line indicates that this line describes a cut Pomeron.}
\label{CorGen}
\end{figure}

\begin{figure}
\begin{minipage}{11cm}
\begin{center}
\includegraphics[width=0.90\textwidth]{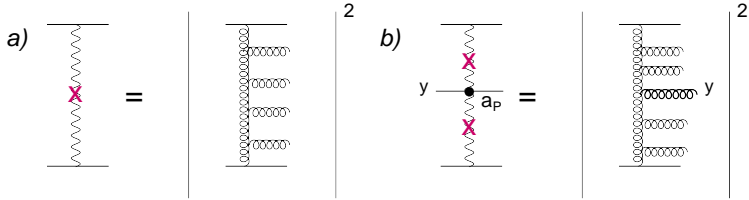}
\end{center}\end{minipage}
\begin{minipage}{6.5cm}
\caption{Shows the main ingredients of \protect\fig{CorGen}: the cut 
Pomeron 
that describes the process
of multiparticle (multigluon) production (\protect\fig{CorCutP}-a)
and the single inclusive  production 
 from the cut Pomeron 
(\protect\fig{CorCutP}-b).} \label{CorCutP} 
\end{minipage}
\end{figure}

It is well known\cite{SOFT} that the most appropriate framework 
to discuss the inclusive processes has
been developed by A.H. Mueller\cite{MUDI} (Mueller diagrams). 
In \fig{CorGen} we show the most general Mueller diagram for 
the double inclusive cross section 
(see also \fig{CorCutP}).  From \fig{CorGen}-a one can see 
that it is necessary  to calculate the amplitudes of
the cut Pomeron interaction with the hadrons, denoted by 
$N\Lb Y - y_1, Y - y_2\Rb$ and $N\Lb y_1, y_2\Rb $.
The fact that we can reduce the calculation of the 
double inclusive production to an evaluation of
$N\Lb Y - y_1, Y - y_2\Rb $  and $ N\Lb y_1, y_2\Rb$ 
stems from the AGK cutting rules\cite{AGK}
which state that the exchanges of the Pomerons from the top to the 
bottom of the Mueller diagram 
cancel each other leading to the general structure of \fig{CorGen}-a. 
 Recall that
the AGK cutting rules are violated in QCD due to the emission diagrams 
from the triple
Pomeron vertex (see \fig{CorGen}-b) (see Ref. \cite{AGKQCD}). 
In our treatment we neglect 
such a violation since $\Gamma\Lb Y - y_1\Rb$ 
turns out to be smaller at high energy than $N\Lb Y - y_1, Y - 
y_2\Rb $ .
Indeed, in the first approximation  
$\Gamma\Lb Y - y_1\Rb  \propto g e^{ \Delta (Y - y_1)}$
while $N\Lb Y - y_1, Y - y_2\Rb\,\propto\, 
\Big(g e^{ \Delta (Y - y_1)}\Big)^2$ and
$\Gamma\Lb Y - y_1\Rb \Big{/} N\Lb Y - y_1, Y - y_2\Rb \,\,\to\,\,0$ 
at large values of $Y - y_1$.
Analyzing the diagrams one can see that their contributions are 
proportional to two parameters which are large at high energy:
\beq \label{LAT}
L\Lb Y \Rb \,\,=\,\,g \Lb b \Rb \frac{G_{3\pom}}{\Delta}\,
e^{ \Delta\,Y}\,;\,\,\,\,\,\,\,\,\mbox{and}
\,\,\,\,\,\,\,\,T\Lb Y \Rb\,\,=\,\,\frac{G^2_{3 \pom}}{\Delta^2}\,
e^{\Delta\,Y}.
\eeq
Note that $\Delta$ in the dominator stems from the integration 
over internal rapidities of the triple Pomeron vertices.
\begin{figure}[h]
\begin{minipage}{11cm}
\begin{center}
\includegraphics[width=0.90\textwidth]{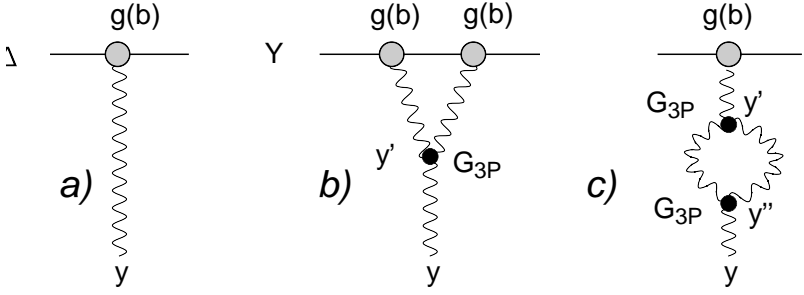}
\end{center}
\end{minipage}
\begin{minipage}{6.5cm}
\caption{ Low order diagrams for $\Gamma\Lb Y - y\Rb $ 
(see \protect\fig{CorGen}). Wavy lines denote the Pomerons.}
\label{CorExDi}
\end{minipage}
\end{figure}
We consider the first three diagrams (see \fig{CorExDi}) 
for $\Gamma\Lb Y - y\Rb$ (see \fig{CorGen}) to illustrate how these two 
parameters appear in the calculations. 
For the diagrams of \fig{CorExDi}-a, \fig{CorExDi}-b and \fig{CorExDi}-c  
we have, respectively,
\bea 
A\Lb \mbox{Pomeron}\Rb  \,\,&=&\,\,g\Lb b \Rb e^{\Delta (Y - y)}; 
\label{DI1}\\
A\Lb \mbox{'fan' diagram}\Rb\,\,&=&\,\, - g^2\Lb b \Rb\,G_{3\pom} 
\int^Y_0 d y' e^{2 \Delta (Y - y')}\,e^{ \Delta  y'}\,\nn\\
&=&\, - g\Lb b \Rb e^{\Delta (Y - y)}\Big(L\Lb Y - y\Rb \,-\, 
g\Lb b \Rb\,\frac{G_{3\pom}}{\Delta}\Big)\,\nn\\
&\xrightarrow{ Y- y \gg 1}& \,- A\Lb \mbox{Pomeron}\Rb\,L\Lb Y - y \Rb; 
\label{DI2}\\
A\Lb \mbox{enhanced diagram}\Rb\,\, & = & \,\,-g^2\Lb b \Rb\,G^2_{3\pom} 
\int^Y_0 d y' \int ^{y'}_0 d y''\, e^{ \Delta (Y - y')} 
e^{2 \Delta (y' - y'')}\,e^{ \Delta  y''}\,\nn\\
&=&\, -g\Lb b \Rb e^{\Delta (Y - y)}\Big(T\Lb Y - y\Rb \,-\, g\Lb b \Rb\,
\frac{G^2_{3\pom}}{\Delta^2}\,(1 + \Delta(Y - y)\Big)\nn\\
& \xrightarrow{ Y- y \gg 1}& \,- A\Lb \mbox{Pomeron}\Rb\,T\Lb Y - y \Rb; 
\label{DI3} \eea
\begin{figure}[h]
\begin{minipage}{11cm}
\begin{center}
\includegraphics[width=0.90\textwidth]{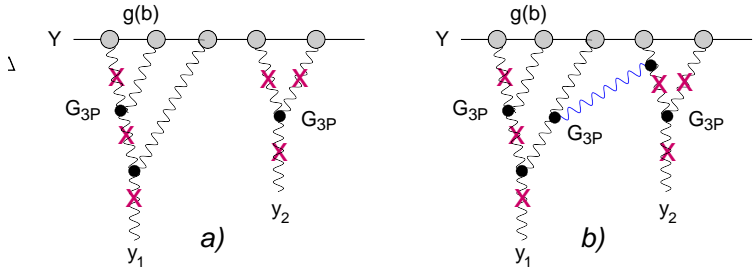}
\end{center}
\end{minipage}
\begin{minipage}{6.5cm}
\caption{The main diagrams  relating to $L\Lb Y - y_i\Rb$ 
(see text), that contribute to the function $N\Lb Y - y_1, Y- y_2\Rb$ 
(\protect\fig{CorMain}-a); and the first diagram with correction that is 
proportional to $T (Y - y_i)$ (\protect\fig{CorMain}-b).
Wavy lines denote the Pomerons. The cross on the wavy line indicates that 
this line describes the cut Pomeron.}
\label{CorMain}
\end{minipage}
\end{figure}
At high energy both $L\Lb Y \Rb \,\gg \,1$ and $T\Lb Y\Rb\,\gg\,1$ 
and we can neglect other contributions in each diagram. 
In this kinematic region each Pomeron diagram is proportional to powers 
of $L\Lb Y\Rb$ and $T\Lb Y \Rb$. 
Therefore, the first approximation is to sum the largest contributions 
at high energies in every Pomeron diagram. 
Such an approach to high energy scattering was proposed by 
Mueller, Patel, Salam and Iancu (MPSI approximation\cite{MPSI}).  
It turns out that the value of $G_{3\pom}$ is rather small 
(see discussion below). Based on this fact we  propose that the 
leading approximation shall be to sum all contributions proportional 
to $L^n\Lb Y - y\Rb$ having in mind the following kinematic region:
\beq \label{KINR}
L \Lb Y - y \Rb\,\,\geq\,1\,;\,\,\,\,T\Lb Y - y \Rb 
\,\,\ll\,\,1;\,\, g(b)\,\ll \,1;\,\,\,G_{3\pom} \,\ll \,1.
\eeq
For the scattering with nuclei $g\Lb b \Rb \propto A^{1/3}$, 
and in this region which covers all reasonable energies,   
the main contribution emanates from 'fan' diagrams 
(see \fig{CorMain} and \fig{CorExDi}- b for the first diagram of this kind). 
The expression for $\Gamma\Lb Y - y\Rb$ is known \cite{SCH,BOKA}:
\beq
\Gamma\Lb Y - y\Rb \,\,=
\,\,\frac{2 \,g\,e^{\Delta (Y - y)}}{1 + L\Lb Y - y\Rb};\label{CAPO}.
\eeq
As we shall see below the factor 2 stems from the initial cut Pomeron.
 Below  we shall obtain   these expressions using a more general technique 
in 
which  we   find the sum of the diagrams in a more general 
kinematic region: 
\beq \label{KINRG}
L \Lb Y - y \Rb\,\,\geq\,1\,;\,\,\,\,~~T\Lb Y - y \Rb 
\,\,\geq\,\,1;\,\,~~~ g(b)\,\ll \,1;\,\,~~\,\,G_{3\pom} \,\ll \,1,
\eeq
selecting contributions of the order of 
$L^m\Lb Y - y\Rb\,T^{n - m} \Lb Y - y \Rb$.
i.e. we shall find the scattering amplitude in the kinematic 
region of \eq{KINRG} using MPSI approximation.

The most important diagrams for $N\Lb Y - y_1,Y - y_2\Rb$ 
are shown in \fig{CorExDi}-a. One can see that the kinematic region of 
\eq{KINR}  $N\Lb Y - y_1,Y - y_2\Rb$ is:
\beq \label{NINLA}
N\Lb Y - y_1,Y - y_2\Rb\,\,=\,\,\Gamma\Lb Y - y_1\Rb\,\Gamma\Lb Y - y_2\Rb.
\eeq
\subsection{Generating function approach}
We believe that the method of a generating function (functional) is 
the most appropriate  method for summing Pomeron diagrams.
In the MPSI approach, one can explicitly see the conservation 
of probability (unitarity constraints) in each step of the evolution in 
rapidity.  
This method was proposed by Mueller in Ref.\cite{MUCD} 
and has been developed in a number of publications(see Ref.\cite{LL} 
and references therein). In Ref.\cite{LEPRY} it was generalized to 
account for the contribution to the inelastic processes
by summing both cut and uncut Pomeron contributions. For completeness of 
the presentation, in this section we shall discuss the main features of 
this method, referring to Refs.\cite{LEPRY, GLM1,GLM2} for  
essential details. 
Following Ref.\cite{LEPRY}, we introduce the generating function 
\beq\label{GENFU}
Z(w, \bar{w},v|Y)=\sum_{k=0}\sum_{l=0}
\sum_{m=0}P(k,l,m|Y) w^k \bar{w}^l v^m,
\eeq 
where, $P(k,l,m|Y)$ stands for the probability to find $k$ 
uncut Pomerons in the amplitude, $l$
uncut Pomerons in the conjugate amplitude 
and $m$ cut Pomerons at some rapidity $Y$. 
$w, \bar{w}$ and $v$ are independent variables. Restricting ourselves by 
taking into account only a Pomeron splitting into two Pomerons, 
we can write the following simple evolution equation: 
\beq \label{ZWWV}
\frac{\partial Z}{\partial Y}\,\, =
\,\, -\,\Delta\Big\{w(1-w)\frac{\partial Z}{\partial w}
- \bar{w}(1-\bar{w})\frac{\partial Z}{\partial \bar{w}} \Big\}
\,\,-\,\, \Delta\Big\{2w\bar{w}-2wv-2\bar{w}v+v^2+v)
\frac{\partial Z}{\partial v}\Big\}.
\eeq

\begin{figure}[h]
\begin{minipage}{11cm}
\begin{center}
\includegraphics[width=0.90\textwidth]{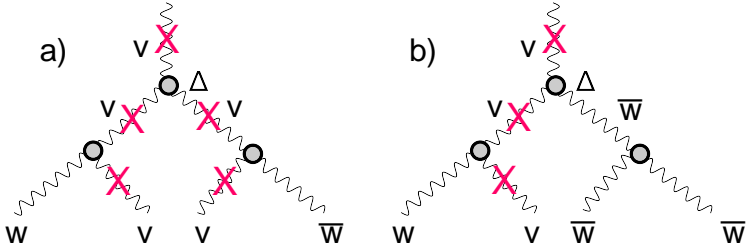}
\end{center}
\end{minipage}
\begin{minipage}{6.5cm}
\caption{Two examples for two steps of evolution in rapidity for the  
generating function $Z\Lb w,\bar{w}, v ; Y\Rb$.
Wavy lines denote the Pomerons. The cross on the wavy line indicates that 
this line describes the cut Pomeron.}
\label{CorVWW}
\end{minipage}
\end{figure}

\fig{CorVWW} illustrates the two steps of evolution in rapidity for 
$Z\Lb w,\bar{w}, v ; Y\Rb$.
The general solution to \eq{ZWWV} has the form
\beq \label{SOLZWWV}
C_1\,Z\Lb w \Rb\,\,+\,C_1 \,Z\Lb \bar{w}\Rb\,+
\,C_2 \,Z\Lb w \,+ \,\bar{w}\,-\, v\Rb,
\eeq
where, $C_1$ and $C_2$ are constants and $Z\Lb \xi\Rb$ 
is the solution to the equation:
\beq \label{ZXI}
\frac{\partial Z}{\partial Y}\,=\,- \,\Delta \xi(1 - \xi)\,
\frac{\partial Z}{\partial \xi}.
\eeq
The particular form of $Z$ and the values of $C_i$ are determined by the 
initial condition at $Y=0$.
\subsection{Amplitude in the MPSI approach: general formula}
The general formula for the amplitude in the MPSI approach has the form 
(see Ref.\cite{LEPRY})
\bea \label{FOMPSI}  
N^{MPSI}\Lb  \gamma, \gamma_{in}|Y\Rb&=&\Big(
\exp \left\{\,-\,\gamma\,\frac{\partial}
{\partial \gamma^{(1)} }\,\frac{\partial}{\partial \gamma^{(2)}}
\,\,-\,\,\gamma\,\frac{\partial}{\partial \bar{\gamma}^{(1)}}\,
\frac{\partial}{\partial \bar{\gamma}^{(2)}}\,
\,+\,\,\gamma_{in}\,\frac{\partial}{\partial \gamma^{(1)}_{in}}\,
\frac{\partial}{\partial \gamma^{(2)}_{in}}\,\right\}
\,-\,1\Big)\nn\\
& &Z\Lb  \gamma^{(1)},\bar{\gamma}^{(1)},\gamma^{(1)}_{in}|Y - Y'\Rb
Z\Lb  \gamma^{(2)},\bar{\gamma}^{(2)},\gamma^{(2)}_{in}|Y'\Rb|_{\gamma^{(i)}\,
=\bar{\gamma}^{(i)} = \gamma^{(i)}_{in}=\,0}, 
\eea
where, $w\,=\, 1 - \gamma$, $\bar{w}\,=\,1 - \bar{\gamma}$ 
and $ v\,=\,1 - \gamma_{in}$.
\begin{figure}[h]
\begin{minipage}{11cm}
\begin{center}
\includegraphics[width=0.85\textwidth]{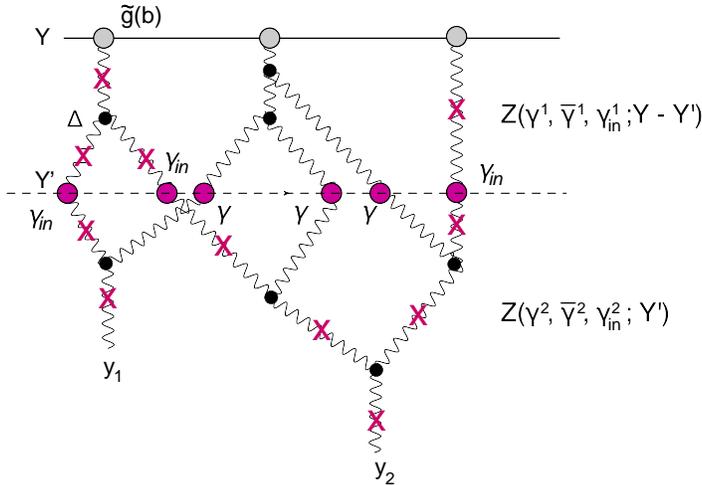}
\end{center}
\end{minipage}
\begin{minipage}{6.5cm}
\caption{An example of  diagrams that contribute to the function 
$N\Lb Y - y_1, Y- y_2\Rb$ 
(see \protect\fig{CorGen}). Wavy lines denote the
Pomerons. The cross on the wavy line indicates that this line describes 
a cut Pomeron.  $\gamma$ is the
amplitude of the dipole-dipole interaction at low energies . The particular
set of diagrams shown in this figure, corresponds to the MPSI
approach \cite{MPSI}.}
\label{CorMPSI}
\end{minipage}
\end{figure}

\eq{FOMPSI} has a very 
simple meaning which is clear from \fig{CorMPSI}. 
The derivatives of the generating functional $Z\Lb  
\gamma^{(1)},\bar{\gamma}^{(1)},\gamma^{(1)}_{in}|Y - Y'\Rb$ 
determine the probability to have cut and uncut Pomerons at 
$Y = Y'$, while the derivatives of 
$Z\Lb \gamma^{(2)},\bar{\gamma}^{(2)},\gamma^{(2)}_{in}|Y'\Rb$ 
lead to the probabilities of the creation of cut and uncut Pomerons from 
two initial cut Pomerons at rapidity $Y'$. 
Two uncut Pomerons interact with the amplitude $\gamma$ 
at rapidity $Y'$ and with the amplitude $\gamma_{in}$ 
in the case of cut Pomerons. 
The phases of the amplitude are given by related signs in \eq{FOMPSI}: 
minus for $\gamma$ and plus for $\gamma_{in}$. 
In addition, we assume that the low energy at which 
the wee partons from two Pomerons interact is large enough to assume that 
$\gamma$ and $\gamma_{in} $ are purely imaginary. 
We denote the  imaginary part of the amplitude,
 by $\gamma$'s.
It follows from the AGK cutting rules that
\beq \label{GAGAIN}
\gamma_{in}\,\,=\,\, 2 \,\gamma.
\eeq

According to \eq{FOMPSI}, the contribution to the scattering 
amplitude of one Pomeron exchange is equal to
\footnote{We suppress the notation of the impact parameter, 
which if needed can be easily be replaced.}
\beq \label{OPEX}
\tilde{g}\,e^{\Delta (Y - Y')}\,\gamma \,e^{\Delta Y'} \,\tilde{g}.
\eeq
For the first 'fan' diagram, \eq{FOMPSI} leads to the following 
contribution:
\beq \label{OFDEX}
\tilde{g} \int^Y_{Y'} d y'  e^{\Delta (Y - y')}\,\Delta \,e^{2\,\Delta 
(y' - Y')} \,\gamma^2\,,e^{2\,\Delta (Y')}\tilde{g}^2,
\eeq
while the first enhanced diagram can be written as
\beq \label{EDEX}
\tilde{g} \int^Y_{Y'} d y'\, e^{\Delta (Y - y')}\,\Delta \, 
\,e^{2\,\Delta (y' - Y')}\,\gamma^2\,\,\int^{Y'}_0 \,d y''\, 
e^{2 \,\Delta (Y' - y'')}\,\Delta
\,e^{\Delta y''} \,\tilde{g}.
\eeq
Comparing these expressions with the Pomeron diagrams 
(see \eq{DI1},\eq{DI2} and \eq{DI3}), we have the correspondence 
between these two approaches,
\beq \label{GLO}
\tilde{g}\,\,=\,\,g/\sqrt{\gamma}\,;\,\,\,\,\,\,\gamma\,\,
=\,\,\frac{G^2_{3\pom}}{\Delta^2}.
\eeq
\subsection{MPSI approximation: instructive examples}
\subsubsection{Glauber-Gribov formula}
\begin{figure}[h]
\begin{tabular}{c c}
\includegraphics[width=0.60\textwidth]{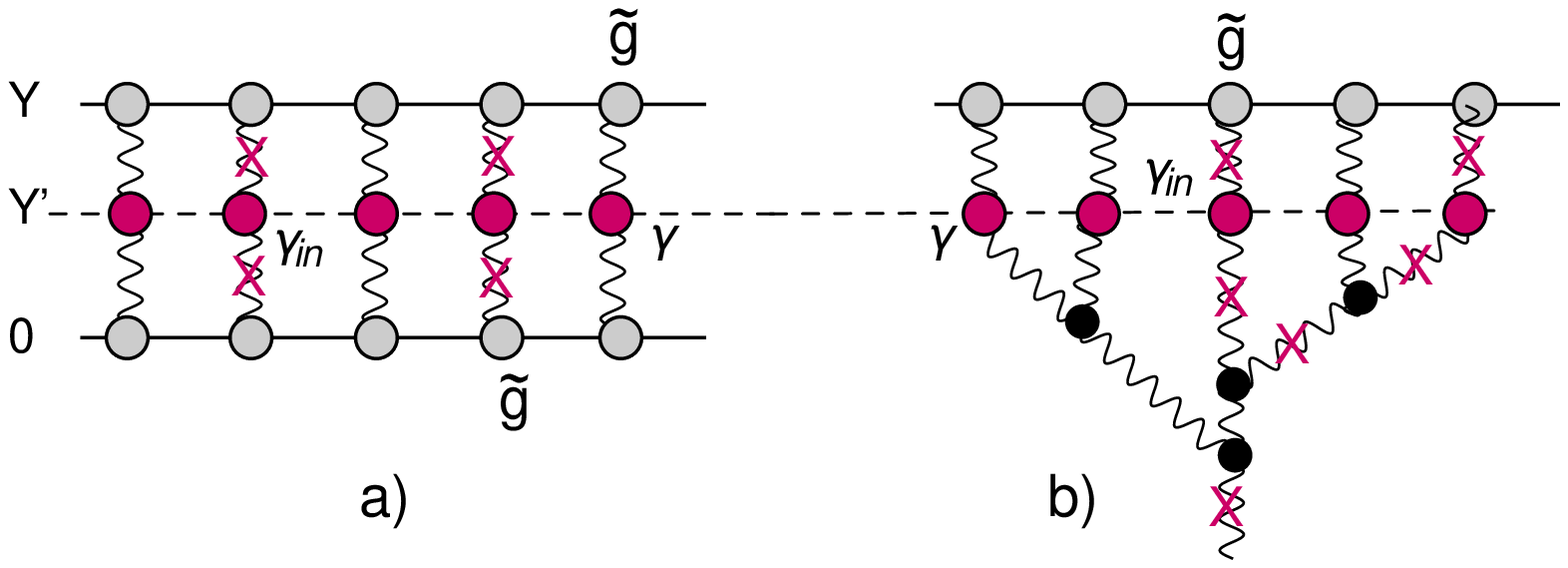}& 
\includegraphics[width=0.35\textwidth]{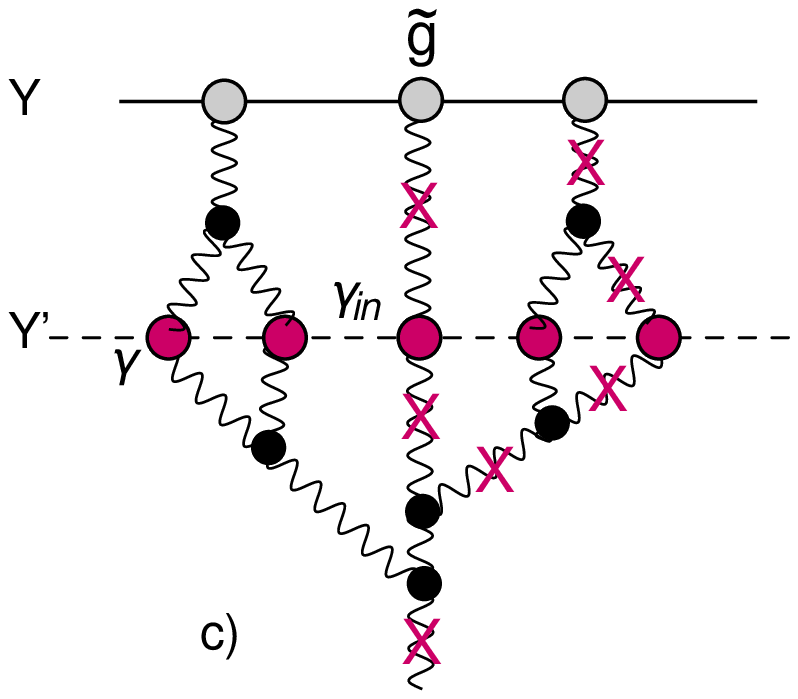}
\end{tabular}
\caption{MPSI approximation: Glauber-Gribov rescattering 
(\protect\fig{CorExMPSI}-a), summation of 'fan' diagrams
 (\protect\fig{CorExMPSI}-b) and  the
 diagrams for single inclusive cross section
 (\protect\fig{CorExMPSI}-c)  Wavy  lines denote the Pomerons.
 The cross on the wavy line  indicates that this line describes 
the cut Pomeron.  $\gamma$ is the
amplitude of the dipole-dipole interaction at low energies.}
\label{CorExMPSI}
\end{figure}

The pattern of calculation of Glauber-Gribov rescatterings due to 
Pomeron exchanges is shown in \fig{CorExMPSI}-a. 
The forms of the generating functions 
$Z\Lb \gamma^{(1)},\bar{\gamma}^{(1)},\gamma^{(1)}_{in}|Y - Y'\Rb$ and 
$ Z\Lb  \gamma^{(2)},\bar{\gamma}^{(2)},\gamma^{(2)}_{in}|Y'\Rb $ 
are simple, 
\bea
Z\Lb  \gamma^{(1)},\bar{\gamma}^{(1)},\gamma^{(1)}_{in}|Y - Y'\Rb &=& 
e^{\tilde{g} \,e^{\Delta(Y - Y')}\,\Lb w^{(1)} 
+ \bar{w}^{(1)} - v^{(1)} - 1\Rb} \,\,
=\,\,e^{\tilde{g} \,e^{\Delta(Y - Y')}\, \Lb\gamma^{(1)} \,
+\,\bar{\gamma}^{(1)} \,-\,\gamma^{(1)}_{in}\Rb} ;\label{ZG1}\\ 
Z\Lb \gamma^{(2)},\bar{\gamma}^{(2)},\gamma^{(2)}_{in}|Y'\Rb    
&=& e^{ \tilde{g} \,e^{\Delta(Y')}\,\Lb 
w^{(2)} + \bar{w}^{(2)} - v^{(2)} - 3\Rb} \,\,=\,\,e^{\tilde{g} 
\,e^{\Delta(Y - Y')}\, \Lb\gamma^{(2)} \,+\,\bar{\gamma}^{(2)} \,
-\,\gamma^{(2)}_{in}\Rb}. \label{ZG2}
\eea
 These generating functions describe the independent 
(without correlations) interaction of  Pomerons with the target and the 
projectile. In the case of nuclei, Pomerons interact with different 
nucleons 
in the nucleus, and the correlations between nucleons in the 
wave function of the nucleus are neglected. 
Note that \eq{FOMPSI} with $Z$'s from \eq{ZG1} and \eq{ZG2} 
do not depend on the sign of $v$ ($\gamma_{in}$). 
However, we shall see below that the choice of the above equation 
is correct since it reproduces \eq{CAPO}, which has been derived 
by summing the Pomeron diagrams.
 
Using \eq{FOMPSI}, we can calculate the inelastic cross section requiring 
that at rapidity $Y'$ we have at least one cut Pomeron 
(one $\gamma_{in}$). 
The  result is:
\beq \label{GGIN}
\sigma_{in}\,\,=\,\,1\,\,-\,\,e^{ -\gamma_{in} \tilde{g}^2\,e^{\Delta Y}}\,\,
=\,\, 1 \,\,-\,\,e^{ - 2\,g^2\,e^{\Delta Y}}. 
\eeq
which reproduces the well known expression for the inelastic cross section 
in 
the Glauber-Gribov approach.
 
We can also calculate the contribution which has no cut Pomeron at 
rapidity $Y'$ (elastic cross sections).
It has the form
\beq \label{GGEL}
\sigma_{el}\,\,=\,\,\Big(1\,\,-\,\,e^{  -\gamma \tilde{g}^2\,
e^{\Delta Y}} \Big)\,\,\Big(1\,\,-\,\,e^{  -\bar{\gamma }\tilde{g}^2\,
e^{\Delta Y}}\Big)
\,\,=\,\, \Big(1 \,\,-\,\,e^{ - g^2\,e^{\Delta Y}}  \Big)^2.
\eeq
The  total cross section is given by: 
\beq \label{GGTXS}
\sigma_{tot} \,=\, \sigma_{el} \,+\, \sigma_{in}\,=
\,2 \Big(1 \,\,-\,\,e^{ - \,g^2\,e^{\Delta Y}}\Big).
\eeq
\subsubsection{Summing 'fan' diagrams}
As one can see from \fig{CorExMPSI}-b, the form of $Z\Lb  
\gamma^{(1)},\bar{\gamma}^{(1)},\gamma^{(1)}_{in}|Y - Y'\Rb$
is the same as in the previous problem. It is given by \eq{ZG1}. 
 To  obtain an expression for $Z\Lb \gamma^{(2)},
\bar{\gamma}^{(2)},\gamma^{(2)}_{in}|Y'\Rb$, we 
need to find $Z$'s and $C_i$ in \eq{SOLZWWV} with the initial condition
\beq \label{IC1}
Z\Lb \gamma^{(2)},\bar{\gamma}^{(2)},\gamma^{(2)}_{in}|Y'= 0\Rb\,\,=\,\,v.
\eeq
 
The resulting solution is of the form (see more details in 
Ref.\cite{LEPRY})
\bea
&&Z\Lb  w,\bar{w},v ; Y' \Rb\,\,=\,\, Z_{el} \Lb w, 
\bar{w}; Y'\Rb\,\,+\,\,Z_{in}\Lb  w,\bar{w},v ; Y' \Rb;\label{SOLSUMZ}\\
&&Z_{el} \Lb w, \bar{w}; Y'\Rb = \frac{ w \, e^{-\Delta  Y'}}
{1+w(e^{- \Delta  Y'}-1)}\,\,+\,\,
\frac{\bar{w}\,e^{- \Delta Y'}}{1+\bar{w} (e^{-\Delta Y'}-1)}\,\,
-\,\,\frac{(w + \bar{w}) e^{-\Delta  Y'}}
{1+(w + \bar{w})(e^{-\Delta\,\, Y'}-1)}; \label{SOLELZ}\\
&& Z_{in}\Lb  w,\bar{w},v ; Y'\Rb \,\,=\,\,\,\frac{(w + \bar{w}) 
e^{-\Delta  Y'}}{1+(w + \bar{w})(e^{-\Delta\,\, Y'}-1)} \,\,-\,\,
\frac{(w + \bar{w}\,-\,v ) e^{-\Delta Y'}}
{1+(w + \bar{w}\,-\,v)(e^{-\Delta\,\, Y'}-1)}. \label{SOLINZ}
\eea 
 
Substituting for $Z_{in}$ in \eq{FOMPSI} we obtain 
for the inelastic part of $\Gamma\Lb Y - y\Rb$ (see \fig{CorGen}-b),
\beq \label{ZIN}
\Gamma_{in}\Lb Y - y\Rb\,\,=\,\,\frac{2\,\tilde{g} \gamma e^{\Delta (Y - y)} }{1 \,\,+\,\,2\,\tilde{g}\gamma e^{ \Delta (Y - y)}}\,\,=\,\,
\frac{2 L\Lb Y - y\Rb}{1 +2 L\Lb Y - y\Rb}.
\eeq
\eq{ZIN} has been derived  from the direct summation of the Pomeron 
diagrams in Ref.\cite{BOKA}.  
  The fact that we reproduce the results 
of Ref.\cite{BOKA} ,  vindicates our choice of 
the generating functions in \eq{ZG1} and \eq{ZG2}.

Using $Z_{el}$ we obtain the elastic contribution which is intimately 
related to the processes of diffraction production:
\beq \label{ZEL}
\Gamma_{el}\Lb Y - y\Rb\,\,=\,\,\frac{2\,\tilde{g} 
\gamma e^{\Delta (Y - y)} }{1 \,\,+\,\,\,\tilde{g}\gamma 
e^{\Delta (Y - y)}}\,\,-\,\,\frac{2\,\tilde{g} \gamma 
e^{\Delta (Y - y)} }{1 \,\,+\,\,2\,\tilde{g}\gamma 
e^{ \Delta (Y - y)}} =\,\, \frac{2 L\Lb Y - y\Rb}{1 + L\Lb Y - y\Rb }\,
-\, \frac{2 L\Lb Y - y\Rb}{1 +2 L\Lb Y - y\Rb}.
\eeq
The resulting $\Gamma\Lb Y -y\Rb$ is given by:
\beq \label{ZSUM}
\Gamma\Lb Y - y\Rb\,\,=\,\,\frac{2 L\Lb Y - y\Rb}{1\,+\,L\Lb Y - y\Rb}.
\eeq
Actually \eq{ZSUM} gives the same expression as \eq{CAPO}. 
The difference  in an extra factor, $\sqrt{\gamma}$, 
 stems from the fact that, 
 we need to take $\tilde{g}$ rather than $g$ in  
the vertex for the Pomeron-hadron interaction. 
\subsubsection{Single inclusive production in MPSI approximation}
As one can see from \fig{CorGen}-c,
to evaluate the single inclusive cross section,
 we  need to calculate $\Gamma\Lb Y - y\Rb$. 
We have done so in the previous section, however, we now want
  to take 
into account both $L^n\Lb Y - y \Rb$ and $T^n\Lb Y - y\Rb$ contributions. 
 From \fig{CorExMPSI}-c we see that $Z\Lb  w^{(2)},\bar{w}^{(2)},v^{(2)}; 
Y' \Rb$ has the form given in \eq{SOLSUMZ}. However, in 
$Z\Lb w^{(1)},\bar{w}^{(1)},v^{(1)} ; Y' \Rb$, we need to take into 
account that each Pomeron at $Y - Y'=0$, creates a cascade of Pomerons 
that is described by \eq{ZWWV}. 
In other words, we need to replace $w^{(1)}$, 
$\bar{w}^{(1)}$ and $v^{(1)}$ in \eq{ZG1} by 
\beq 
w^{(1)} \rightarrow  \frac{ w^{(1)} \, 
e^{-\Delta (Y - Y')}}{1+w^{(1)}(e^{- \Delta (Y - Y')}-1)};
\,\,\,\,\,\,\,\,\,\,\,~~~~~~~~~~~~~~~~~~~~~~~~~~~~~~~~~~
\,\,\,\,\bar{w}^{(1)}  \rightarrow  \frac{ \bar{w}^{(1)} 
\, e^{-\Delta (Y -   Y')}}
{1+ \bar{w}^{(1)}(e^{- \Delta (Y - Y')}-1)};\label{SIN12}
\eeq
\beq 
v^{(1)} \rightarrow  \frac{ w^{(1)} \, 
e^{-\Delta (Y -   Y')}}{1+w^{(1)}
(e^{- \Delta (Y - Y')}-1)}\,+\,\frac{ \bar{w}^{(1)} \, 
e^{-\Delta (Y -   Y')}}{1+ \bar{w}^{(1)}(e^{- \Delta (Y - Y')}-1)}
\,-\,\frac{(w^{(1)} + \bar{w}^{(1)}\,-\,v^{(1)} ) 
e^{-\Delta (Y -   Y')}}{1+(w + \bar{w}\,-\,v)
(e^{-\Delta\,\,(Y - Y'}-1)}.\label{SIN3}
\eeq

Using these substitutions we obtain
\beq \label{ZSIN1}
Z\Lb  w^{(1)},\bar{w}^{(1)},v^{(1)} ; Y - Y' \Rb
\,\,\,=\,\,\exp\Lb \frac{(w^{(1)} + \bar{w}^{(1)}\,-\,v^{(1)} ) 
e^{-\Delta (Y -   Y')}}{1+(w + \bar{w}\,-\,v)(e^{-\Delta\,\,(Y - Y'}-1)}\Rb.
\eeq
Using the generating function for Laguerre polynomials
(see Ref.\cite{RY} formula {\bf 8.973(1)}), 
\beq \label{LPGF}
(1 - z)^{- \alpha - 1}\, \exp\Lb \frac{x\,z}{z - 1}\Rb\,\,\,
=\,\,\,\sum^{\infty}_{n = 0}\,L^{\alpha}_n\Lb x \Rb \,z^n.
\eeq
We obtain for \eq{ZSIN1}
\beq \label{ZSER}
Z\Lb  w^{(1)},\bar{w}^{(1)},v^{(1)} ; Y' \Rb\,\,\,=\,\,\,-
\,\sum^\infty_{n=0}\,L^{-1}_n\Lb\tilde{g}_i\Rb\,\Lb - \Lb\gamma^{(1)} \,+
\,\bar{\gamma}^{(1)}\,-\, \gamma^{(1)}_{in}\Rb e^{\Delta (Y - Y')}\Rb^n.
\eeq
 From \eq{FOMPSI} using 
\beq \label{DIF}
\frac{\partial^l}{\partial^l \gamma^{(1)}}\,
\frac{\partial^m}{\partial^m \bar{ \gamma}^{(1)}}\,
\frac{\partial^{n - l - m}}{\partial^{n - l - m} 
\gamma^{(1)}_{in}} \Lb\gamma^{(1)} \,+
\,\bar{\gamma}^{(1)}\,-\, \gamma^{(1)}_{in}\Rb^n \,\,= \Lb - 1\Rb^{n - l - m}\,n!.
\eeq
We obtain 
\beq \label{GAZ}
\Gamma\Lb Y -  y \Rb\,\,=\,\, \sum^\infty_{n=1}\,L^{-1}_n\Lb
\tilde{g}_i\Rb\,n!\,\Lb -  \gamma \,e^{\Delta Y}  
\Rb^n\,\,=\,\, \sum^\infty_{n=1}\,L^{-1}_n\Lb\tilde{g}_i\Rb
\,n! \,\Lb - 1 \Rb^n\,T^n\Lb Y - y\Rb.
\eeq

Introducing $n!\,=\,\int^\infty_0 d \xi \xi^n \,\exp\Lb - \xi\Rb$ 
we reduce \eq{GAZ} to the form
\beq \label{GAAN}
\Gamma\Lb L\Lb Y -  y\Rb, T\Lb Y - y\Rb \Rb\,\,=
\,\,\int^\infty_0 d \xi \, e^{ - \xi}\,\Lb 
e^{-\,\frac{\xi\, \tilde{g}\,\gamma\,
e^{\Delta (Y - y)}}{1 \,+\,\xi\,\gamma\,
e^{\Delta (Y - y)}}}\,\,-\,\,1\Rb\,\,=\,\,\int^\infty_0 d \xi\, 
e^{ - \xi}\Lb 
e^{-\,\frac{ \xi\, L (Y - y)}{1 \,+\,\xi\,T (Y - y)}}\,-\,1\Rb.
\eeq
Using \eq{GAAN} we obtain the following result for the single inclusive 
cross section:
\beq \label{SINFIN}
\frac{d \sigma}{d y}\,\,=\,\,a_\pom\,\Gamma\Lb 
L\Lb Y -  y\Rb, T\Lb Y - y\Rb \Rb\,\Gamma\Lb L\Lb   y\Rb, T\Lb y\Rb \Rb,
\eeq
where, $a_\pom$ denotes the vertex of emission of the hadron from Pomeron 
(see \fig{CorGen} and \fig{CorCutP}-b).
\subsection{The Correlation function in MPSI approximation}
Calculating  $N\Lb Y - y_1,Y - y_2\Rb$ (see \fig{CorGen}-a) we   use 
$ Z\Lb  w^{(1)},\bar{w}^{(1)},v^{(1)} ; Y - Y' \Rb$, given by \eq{ZSIN1}, 
as one can see from \fig{CorMPSI}.
However, $Z\Lb \gamma^{(2)},\bar{\gamma}^{(2)},\gamma^{(2)}_{in}|Y'\Rb$ 
is different from the expression which has been used in the calculation of  
the single inclusive cross section, and it can be written as:
\bea \label{DINZ}
&&Z\Lb\gamma^{(2)},\bar{\gamma}^{(2)},\gamma^{(2)}_{in}|Y'-y_1,Y'-y_2\Rb\,\,=\\
&&~~~~~~~~~~~~~~~~~~~\,\,Z \Lb \mbox{\eq{SOLSUMZ}}| \gamma^{(2)},
\bar{\gamma}^{(2)},\gamma^{(2)}_{in}|Y - y_1\Rb \,Z \Lb 
\mbox{\eq{SOLSUMZ}}| \gamma^{(2)},\bar{\gamma}^{(2)},
\gamma^{(2)}_{in}|Y-y_2\Rb\nn.
\eea
First, we calculate $N\Lb Y - y_, Y - y_2\Rb $ at $y_1=y_2$. Using \eq{ZSER} 
and \eq{DIF} we obtain from \eq{FOMPSI} that
\bea
N\Lb Y - y_1, Y - y_1\Rb\,\,&=&\,\,\sum^\infty_{n=1}\,
L^{-1}_n\Lb\tilde{g}_i\Rb\,\,n! \,(n - 1)\,\Lb - 1 \Rb^n\,T^n\Lb Y - y\Rb
\,\,\label{NDIN1}\\
&=&\,\,T^2\frac{d}{d T} \,(1/T) \Big\{\sum^\infty_{n=1}\,
L^{-1}_n\Lb\tilde{g}_i\Rb\,\,n! \,\Lb - 1 \Rb^n\,T^n\Lb Y - y\Rb\Big\}\nn\\
&=&\int^\infty_0 d \xi \,e^{-\xi}\left\{ 1 \,+ \,
e^{- \frac{\xi L\Lb Y - y_1\Rb}{1\,+\,\xi\,T\Lb Y - y_1\Rb}}\Lb
\frac{- 1  - \xi T\Lb Y - y_1\Rb \,-\,\xi L\Lb Y - y_1\Rb}
{\Lb 1 \,+\,\xi T\Lb Y - y_1\Rb\Rb^2}\Rb\right\}. \label{NDIN2}
\eea
At $L\Lb Y - y_1\Rb \gg 1$ we expand \eq{NDIN2} to estimate the importance 
of the 
the correction depending on $T\Lb Y - y\Rb$. The first four terms are 
given by:
\bea\label{NEXP}
&&N\Lb L\Lb Y - y_1\Rb, T\Lb Y - y_1\Rb;  L\Lb Y - y_1\Rb, T\Lb Y - y_1\Rb \Rb\,\,=\,\,\frac{L^2\Lb Y - y_1\Rb}{ \Lb 1 + L\Lb Y - y_1\Rb\Rb^2}\\
&&\,-\,4 \frac{L^2\Lb Y - y_1\Rb\,T\Lb Y - y_1\Rb}{ \Lb 1 + L\Lb Y - y_1\Rb\Rb^4}\,-\,12  \frac{L^3\Lb Y - y_1\Rb\,T^2\Lb Y - y_1\Rb}{ \Lb 1 + L\Lb Y - y_1\Rb\Rb^6}\,-\, 48   \frac{L^4\Lb Y - y_1\Rb\,T^3\Lb Y - y_1\Rb}{ \Lb 1 + L\Lb Y - y_1\Rb\Rb^8}\,-\,\,\dots\nn\\
&&= \,\frac{L^2\Lb Y - y_1\Rb}{ \Lb 1 + L\Lb Y - y_1\Rb\Rb^2}
\,\,-\,\,2 \sum_{n=2} n!\frac{L^n\Lb Y - y_1\Rb\,T^{n - 1}\Lb Y - y_1\Rb}
{\Lb 1 + L\Lb Y - y_1\Rb\Rb^{2 n}}.
\eea
Note that all corrections have minus signs and the function 
of \eq{NDIN2} gives the analytical summation of the asymptotic series of 
\eq{NEXP}.
For $y_1 \neq y_2$ we have a more complex answer, namely,
\bea \label{NY1Y2}
&&N\Lb L\Lb Y - y_1\Rb, T\Lb Y - y_1\Rb;  
L\Lb Y - y_2\Rb, T\Lb Y - y_2\Rb \Rb\,\,=\\
&& \,\,~~~~~\frac{ \Big\{ T\Lb Y - y_2\Rb  \Gamma\Lb L\Lb Y - y_1\Rb,  
T\Lb Y - y_1\Rb\Rb\,\,-\,\,  
T\Lb Y - y_1\Rb  \Gamma\Lb L\Lb Y - y_2\Rb,  
T\Lb Y - y_2\Rb\Rb\Big\}}{\Lb T\Lb Y - y_2\Rb\,-\,
T\Lb Y - y_1\Rb\Rb}\nn.
\eea

The double inclusive cross section can be written as (see \fig{CorGen}-a)
\bea \label{DINFIN}
&&\frac{d^2 \sigma}{ d y_1\,d y_2}\,\,=\\
&&~~~~~~~~~\,\,a^2_\pom\,N\Lb L\Lb Y - y_1\Rb, T\Lb Y - y_1\Rb;  
L\Lb Y - y_1\Rb, T\Lb Y - y_1\Rb \Rb\,N\Lb L\Lb  y_1\Rb, T\Lb y_1\Rb;  
L\Lb  y_1\Rb, T\Lb  y_1\Rb \Rb\nn.
\eea

\section{Correlations in a model for soft interactions}
Recently  considerable progress has been
achieved in building  models for  soft scattering at high 
energies\cite{GLMLAST,GLM1,GLM2,KAP,KMR,OST}. The main 
ingredient of these 
models is the soft Pomeron with a relatively large  intercept 
$\Delta_{\pom} = \alpha_{\pom} - 1 = 0.2  - 0.4$ 
and exceedingly small slope $\alpha_{\pom}^{\prime} \simeq 0.02\,GeV^{-2}$. 
Such a Pomeron appears in N=4 SYM
\cite{BST,HIM,COCO,BEPI,LMKS} with a large coupling.  This is, at present 
,  is the only theory that allows us
to treat the strong interaction on the theoretical basis. Having 
$\alpha_{\pom}^{\prime} \to 0$, the Pomeron in 
these models has a natural matching with the hard Pomeron that occurs in 
 perturbative QCD. Therefore,
these models could be a first step in building a selfconsistent 
theoretical description of the soft interaction at high energy, 
in spite of its many phenomenological parameters (of the order of 10-15)  
in every model.

In this section we shall discuss the size of the correlation function in 
our model\cite{GLMLAST,GLM1,GLM2}. This model describes the LHC data  
(see Refs.\cite{ALICE,ATLAS,CMS,TOTEM}), including the single inclusive 
cross section. 
 Thus our next step is to try,  to understand the predicted 
size of the long range rapidity correlations in this model.
\subsection{Estimates of the rapidity correlation function}
In Table 1 we present the main parameters of our model. The  
parameter $T\Lb Y\Rb = \gamma e^{\Delta_\pom Y} $ is small in our model 
reaching about 0.3 at the LHC energies. 
However, $L_i\Lb Y; b \Rb = g_i(b\,)G_{3 \pom}/\Delta_\pom 
e^{\Delta_\pom Y}$ is large (see Ref.\cite{GLMLAST, GLM1}). 
\beq \label{GB}
g_i\Lb b \Rb\,\,=\,\,g_i\,S_i(b)\,=\,\frac{g_i}{4 \pi}\,m^3_i \,b\,
K_1\Lb m_i b\Rb.
\eeq
One can see that $ L_2\Lb Y, b=0\Rb$ is as large as 25 at $Y = 17.7$. 
Therefore, we can evaluate the influence of 
the corrections with respect to $T\Lb Y\Rb$, by calculating the 
contributions 
of two diagrams: \fig{CorMain}-a (the main contribution) 
and \fig{CorMain} - b (the corrections $\propto T\Lb Y\Rb$).
\TABLE[ht]{
\begin{tabular}{||l|l|l|l||}
\hline\hline
$\Delta_\pom $ & $\beta$ & $g_1\,(GeV^{-1})$ &  $g_2\,(GeV^{-)}$\\ \hline
0.23 & 0.46   & 1.89   &
61.99 \\ \hline  \hline
 $m_1$ \,(GeV)&
$m_2$\,(GeV)  & $\gamma $ &   $G_{3\pom}/\Delta_\pom\,(GeV^{-1})$ \\ \hline
 5 &1.71& 0.0045 & 0.03 \\ \hline\hline
\end{tabular}
\caption{Fitted parameters for our model. $\alpha'_\pom = 0.028 
\,GeV^{-2}.$)}
\label{t1}}
We need to use the first two terms of \eq{NEXP} to calculate 
$N\Lb L\Lb Y - y_1\Rb, T\Lb Y - y_1\Rb;  L\Lb Y - y_1\Rb, T\Lb Y - y_1\Rb \Rb$ 
 while being  careful  to account for the correct $b$ dependence.

Introducing two functions,
\beq \label{FUG}
\Ga^{(1)}\Lb L_i\Lb Y - y; b\Rb\Rb\,\,= \,\,\Delta_\pom\,
\frac{ L_i\Lb Y - y; b\Rb}{ 1 \,+\, L_i\Lb Y - y; b\Rb}\,;~~~
\Ga^{(2)}\Lb L_i\Lb Y - y; b\Rb\Rb\,\,= \,\,\Delta_\pom\,
\frac{ L_i\Lb Y - y; b\Rb}{\Lb 1 \,+\, L_i\Lb Y - y; b\Rb\Rb^2}.
\eeq
We can see that \fig{CorMain}-a has the following contributions:
\bea \label{MAINSIN}
\frac{d ^2 \sigma^{(0)}}{d y_1\, d y_2}\,\,&=&\,\,\int d^2 b 
\Big\{\int d^2 b'\,  \Ga^{(1)}\Lb L_i\Lb Y - y_1; \vec{b}'\Rb\Rb 
\Ga^{(1)}\Lb L_i\Lb y_1; \vec{b} - \vec{b}'\Rb\Rb \Big\}\nn\\
&\times &\,\Big\{   \int d^2 b'\,  \Ga^{(1)}\Lb L_i\Lb Y - y_1; 
\vec{b}'\Rb\Rb   \Ga^{(1)}\Lb L_i\Lb y_2; \vec{b} - \vec{b}'\Rb\Rb \Big\},  
\eea
while for \fig{CorMain}-b we have, for $y_1 >y_2$:
\bea \label{FIRSTIN}
&& \frac{d ^2 \sigma^{(1)}}{d y_1\, d y_2}\,\,=\,\, - \, 4 T\Lb Y - y_1\Rb\\
&&\,\,\times \int d^2 b d^2 b'  \Ga^{(2)}\Lb L_i\Lb Y - y_1; 
\vec{b} - \vec{b}'\Rb\Rb\,   \Ga^{(2)}\Lb L_i\Lb Y - y_2; 
\vec{b} - \vec{b}'\Rb\Rb \Ga^{(1)}\Lb L_i\Lb  y_1; \vec{b}'\Rb \Rb  
\Ga^{(1)}\Lb L_i\Lb y_2; \vec{b}'\Rb\Rb \nn
\eea 

Performing the calculations, we found that 
the correlation function $ R\Lb y_1= Y/2, y_2=Y/2\Rb$ (see \eq{I1}) 
is equal to $ R^{(0)}\Lb y_1= Y/2,y_2=Y/2\Rb = 13$ at the Tevatron energy 
and  $R^{(0)}\Lb y_1= Y/2, y_2=Y/2\Rb = 16$ at $W = 7\,TeV$. 
The corrections turn out to be small ($< 5\%$) for both energies. 
Indeed, large correlations were not seen at Tevatron. 
\subsection{Improvement of the model}
\begin{figure}
\begin{center}
\includegraphics[width=0.7\textwidth]{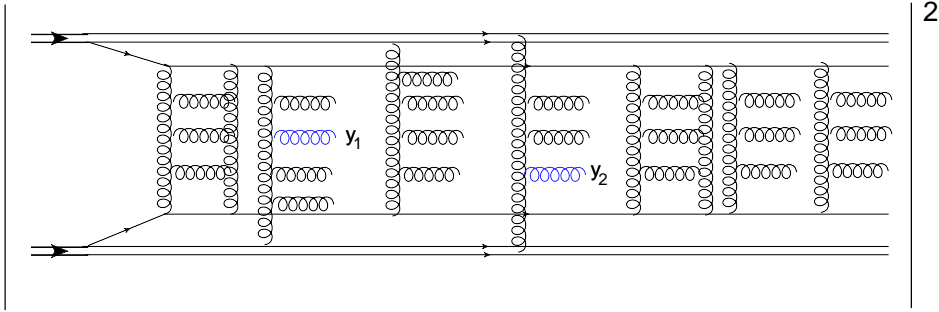}
\end{center}
\caption{ The general diagram for the production of two hadrons(gluons) 
with rapidities $y_1$ and $y_2$.}
\label{Cor2had}
\end{figure}

 \eq{MAINSIN} is written without taking into account any corrections due 
to energy conservation. As has been discussed in the 80'th 
(see Refs. \cite{QGSM,DUM}), these corrections  are important 
for the calculation of the correlations. Generally speaking, 
in Pomeron calculus the long range correlations in rapidity stem from the 
production of two hadrons from two different Pomerons 
(two different parton showers, see \fig{Cor2had}). 
In other words, two hadrons in the central rapidity region can be 
produced 
in an event with more than two parton showers (see \fig{Cor2had}). 
 This is shown in \fig{CorExMPSI}-a 
in an eikonal type model, where the proton-proton scattering amplitude is 
written as:
\beq\label{GGAP}
A\Lb s, b \Rb\,\,=\,\,i \Big(1 \,\,\,-\,\,e^{ - \h \Omega\Lb s,b\Rb}\Big).
\eeq
The cross section of $n$ parton showers production is equal to 
(see Refs.\cite{QGSM,DUM} and references therein)
\beq \label{IM1}
\sigma_{n-\mbox{showers}}\,\,\,=\,\,\,\int d^2 b\,\,
\frac{\Omega^n\Lb s, b \Rb }{n!}\,\,e^{ - \Omega\Lb s, b \Rb}.
\eeq 
\eq{IM1} shows that the parton showers are distributed 
according to Poisson distribution with an average 
number of parton showers $\Omega\Lb s, b \Rb$ which has the 
following form in the simple model of \eq{GGAP}:
\beq \label{IMOM}
\Omega\Lb s,b \Rb\,\,=\,\,\int d^2 b' g\Lb b'\Rb \,g\Lb \vec{b}\,
-\,\vec{b}'\Rb\,\Lb \frac{s}{s_0}\Rb^{\Delta_\pom}.
\eeq

\begin{figure}[h]
\begin{center}
\includegraphics[width=0.7\textwidth]{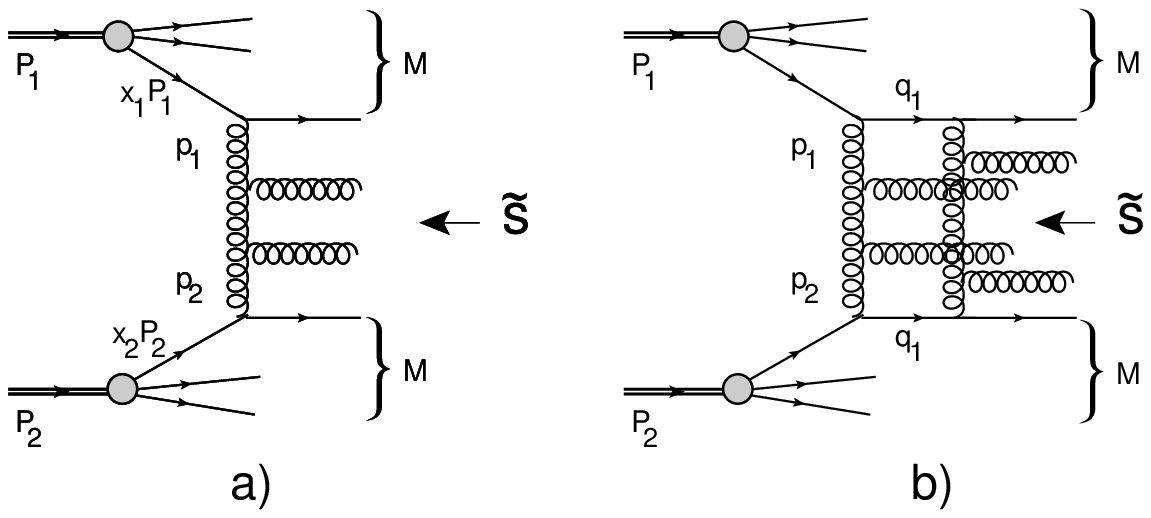}
\end{center}
\caption{ Production of one(\protect\fig{CorKinM}-a) and 
two(\protect\fig{CorKinM}-b) parton showers.}
\label{CorKinM}
\end{figure}

However, the simple \eq{IM1} has to be modified to account for the fact 
that the energy of the parton shower is not equal to  $W = \sqrt{s}$,  
but it is smaller or equal to $\tilde{W}=\sqrt{x_1 x_2 s}$ 
(see \fig{CorKinM}). The easiest way to find $x_1$ and $x_2$ is to assume 
that both $p^2_1 = p^2_2 = - \bar{Q}^2  \gg  \mu^2_{\mbox{soft}}$, where 
$\mu_{\mbox{soft}}$ is the scale of the soft interactions 
$\mu_{\mbox{soft}}\, \sim \,\Lambda_{QCD}$. In Ref.\cite{GLMDM} we 
have argued 
that for a Pomeron $\bar{Q}^2 \approx 2 \,GeV^2 \,\gg \,\mu_{\mbox{soft}}$. 
Bearing this in mind, 
the energy variable $x_1$ ($x_2$) for gluon-hadron scattering 
is equal to 
\bea \label{X}
&&0  \,=\, (x_1\,P_1 + p_1)^2\, =\, -\, \bar{Q}^2 \,
+ \,x _1\,2\, p_1\cdot P_1; ~~~~~~~   p^2_1\, 
=\, -\, \bar{Q}^2; ~~~~~~~~~~~~~~x_1\,
=\,\frac{\bar{Q}^2}{ M^2 + \bar{Q}^2}.
\eea
$p_1$, $P_1$ and $x_1P_1$ are the momenta of the 
gluon, the hadron and the parton (quark or gluon) with which the 
 initial gluon interacts. From \eq{X} one can see that
\beq \label{STILDE}
\tilde{s}\,\,=\,\,x_1 x_2 S \,\,=\,\,\frac{s \,\bar{Q}^4}{M^4}
\eeq
For the second parton shower $\tilde{s} = (q_1 + q_2)^2$ (see 
\fig{CorKinM}), 
where $q_i\,=\,\Big(x^{q_i}_1 \,P_1, x^{q_i}_2 \,P_2 , \,\vec{q}_{i,\perp}
\Big)$. Using the conservation of momentum we see that 
$x^{q_1}_1\,=\,x_1 + x^{g_1}_1$ and $ x^{q_2}_2\,=\,x_2 + x^{g_2}_2$. 
Note that $g_1$ and $g_2$ denote the gluons with momenta $p_1$ and $p_2$ 
respectively (see \fig{CorKinM}). Vectors $p_1$ and $p_2$  take the form:
$p_1\,=\,\Big(x^{g_1}_1 P_1, x^{g_1}_2 P_2 , \,\vec{p}_{1,\perp}\Big)$ 
and $p_2\,=\,\Big(x^{g_2}_1\, P_1, x^{g_2}_2 \,P_2 , 
\,\vec{p}_{1,\perp}\Big)$. Bear in mind the following equations:
\bea \label{IM10}
\hspace{-0.7cm}&&(p_1 +  P_1)^2 \,=\,M^2;~~~~ x^{g_1}_2\,\,=\,\,\frac{M^2}{s};
~~~ p^2_1\,=\,x^{g_1}_1\,,x^{g_1}_2\,s \,+\,p^2_{1,\perp} \,=\,\bar{Q}^2;
~~~ x^{g_1}_1\,<\,\frac{\bar{Q}^2}{x^{g_1}_2\,s}\,=\,\frac{\bar{Q}^2}{M^2}
\,\,\ll\,\,x_1. \nn\\
\hspace{-0.7cm}&&(p_2 +  P_2)^2 \,=\,M^2;~~~~ x^{g_2}_1\,\,=\,\,
\frac{M^2}{s};~~~ p^2_2\,=\,x^{g_2}_2\,,x^{g_2}_1\,s \,+\,p^2_{2,\perp} \
,=\,\bar{Q}^2;~~~ x^{g_2}_2\,<\,\frac{\bar{Q}^2}{x^{g_2}_1\,s}\,=
\,\frac{\bar{Q}^2}{M^2}\,\,\ll\,\,x_2.\eea
                                                 
Therefore, the value of $\tilde{s}$ for the second parton shower turns out 
to be the same as for the first one for $M^2 \,\gg\,\bar{Q}^2$.
The value of $M$ can be estimated using the quark structure function as it 
has been suggested in Ref.\cite{GLMDM}. Indeed,
\beq \label{M}
\langle|M^2|\rangle\,\,=\,\,\frac{\int \frac{d M^2}{M^2}\,M^2 \,
q\Big(\frac{\bar{Q}^2}{M^2 + \bar{Q}^2},\bar{Q}^2\Big)}{\int 
\frac{d M^2}{M^2} \,q\Big(\frac{\bar{Q}^2}{M^2 + 
\bar{Q}^2},\bar{Q}^2\Big)}.
\eeq
Using $\bar{Q}  = 1\,GeV$ and $q\Lb x,\bar{Q}^2\Rb$ given by 
a combined fit\cite{HERAPDF} of  H1 and ZEUS data (HERAPDF01) we obtain 
that $\langle|M^2|\rangle\,\,\approx\,\, 87 \,GeV^2$ which 
is much larger than $\bar{Q}^2$.
                                            
However, the scale of hardness $\bar{Q} $ in CGC/saturation approach is 
proportional to the saturation momentum $Q_s$  ($\bar{Q} \,\propto\,Q_s$) 
and, therefore , depends on energy. Such energy dependence of $\bar{Q}$ 
induces the dependence of average mass $M$ on energy. 
Assuming that $Q_s^2\, \propto \,s^\lambda$ with $\lambda = 0.24$ 
we found that in the energy range $W = 0.9\, \div 7\,TeV$ the typical $M^2 = M^2_0(W=0.9 \,TeV)\,s^\beta$ with $\beta=0.07$.

Taking into account \eq{STILDE} one can re-write \eq{IM1} in the form 
$\langle|M^2|\rangle\,\,=\,\,M^2_0\,\Lb \frac{s}{s_0}\Rb^\beta$ with 
$\beta = 0.14$ and $\sqrt{s}_0 = 0.9\,TeV$ . $M_0 $ is equal to $10\,GeV$ 
.
\beq \label{IM2}
\sigma_{n-\mbox{showers}}\Lb s \Rb\,\,\,=\,\,\,\int d^2 b\,\,
\frac{\Omega^n\Lb \tilde{s}, b \Rb }{n!}\,\,e^{-\Omega\Lb s, b \Rb}.
\eeq

We need to sum over $n \geq 2$ to get the double inclusive production 
cross section,
\beq \label{IM3}
\frac{d^2 \sigma}{d y_1\, d y_2}\,\,\,=\,\,\,2\,a_\pom^2\,\sum_{n = 2}\,
\sigma_{n-\mbox{showers}}\Lb s \Rb\,\,=\,\,
a^2_\pom\,\int d^2 b \,\,\Omega^2\Lb \tilde{s}, b \Rb \,
e^{  \Omega\Lb \tilde{s}, b \Rb\,-\,  \Omega\Lb s, b \Rb },
\eeq                                        
where, $a_\pom$ is a new vertex defined as shown in \fig{CorCutP}-b). The 
factor 2 stems from the possibility to emit a hadron with rapidity $y_1$ 
from each of two parton showers. One can see that the double inclusive cross 
section does not depend on $y_1$ and $y_2$, leading to  the long range 
rapidity correlation.
\subsection{Rapidity long range correlations in GLM  model for 
soft interactions at high energy}
\begin{figure}[h]
\begin{center}
\includegraphics[width=0.7\textwidth]{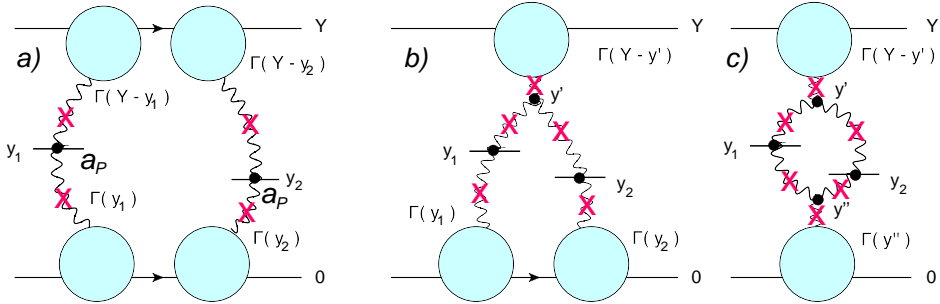}
\end{center}
\caption{Mueller diagrams for double inclusive 
production in the GLM model\cite{GLMLAST,GLM1,GLM2}. 
Crosses mark the cut Pomerons. $\Ga\Lb y \Rb$ is given by \protect\eq{CAPO}. 
All rapidities are in the laboratory reference frame.}
\label{CorGLM}
\end{figure}

In the model for soft interactions that has been suggested in 
Refs.\cite{GLMLAST,GLM1,GLM2} (GLM model) we evaluate more complicated sum 
of  diagrams than in \eq{GGAP}. 
The different contributions to the two particle correlation in 
this model are shown in \fig{CorGLM}. 
\subsubsection{The main ingredients of the GLM model}
{\it  Eikonal diagrams:}\\ 
 In order to account for diffraction dissociation 
in the states with masses  
that are much smaller than
the initial energy, we use the simple two channel 
Good-Walker model. In this model we 
introduce two eigen wave functions, 
$\psi_1$ and $\psi_2$, which
diagonalize the 2x2 interaction matrix ${\bf T}$,
\beq \label{2CHM}
A_{i,k}=<\psi_i\,\psi_k|\mathbf{T}|\psi_{i'}\,\psi_{k'}>=
A_{i,k}\,\delta_{i,i'}\,\delta_{k,k'}.
\eeq                                      
The two observed states are an hadron whose wave function we denote
by $\psi_h$,
and a diffractive state with a wave function $\psi_D$, 
 which is the sum of all the Fock diffractive states. 
These two 
observed states can be written in the form
\beq \label{2CHM31}
\psi_h=\alpha\,\psi_1+\beta\,\psi_2\,,\,\,\,\,\,\,\,\,\,
\psi_D=-\beta\,\psi_1+\alpha \,\psi_2\,,
\eeq
where, $\alpha^2+\beta^2=1$.
For each state we sum the eikonal diagrams of \fig{CorExMPSI}-a 
using \eq{GGAP}. The first contribution to $\Omega\Lb s,b\Rb$ 
is the exchange of a single Pomeron. However, the Pomeron interaction 
leads to a more complicated expression for $\Omega\Lb s, b\Rb$.

{\it Enhanced diagrams:}\\ 
In our model\cite{GLM2}, the Pomeron's Green function which 
includes all enhanced diagrams, is approximated using the MPSI 
procedure\cite{MPSI}, in which a multi Pomeron interaction
(taking into account only triple Pomeron vertices) is          
approximated by large Pomeron loops of rapidity size of $\ln s$. 
We obtain
\beq \label{GFPOM}                  
G_{\pom}\Lb Y\Rb\,\,=\,\,1 \,-\,\\exp\Lb \frac{1}{T\Lb Y\Rb}\Rb\,
\frac{1}{T\Lb Y\Rb}\,\Gamma\Lb 0,\frac{1}{T\Lb Y\Rb}\Rb, 
\eeq
in which:
\beq \label{ES11}
T\Lb Y \Rb\,\,\,=\,\,\gamma\,e^{\Delta_{\pom} Y}.
\eeq
$\Gamma\Lb 0, 1/T\Rb$ is the incomplete gamma function
(see formulae {\bf 8.35} in Ref.\cite{RY}).

{\it Semi-enhanced (net) diagrams:}\\ 
A brief glance at the values of
the parameters of our model (see Ref.\cite{GLMLAST} and Table 1), 
shows that we have a new small parameter, 
$T\Lb Y\Rb \,\,=\,\,G^2_{3 \pom} \Lb s/s_0\Rb^{\Delta_\pom } \,\ll\,1$,
while, $L_i\Lb Y, b \Rb\,\,=\,\,G_{3\pom} \,g_i \Lb b \Rb\,\Lb 
s/s_0\Rb^{\Delta_\pom} \,\approx\, 1$.  
We call the diagrams which are
proportional to $L^n_i\Lb Y, b \Rb$, but do not
contain any of the $T^n\Lb Y, b \Rb$ contributions, net diagrams.
Summing the net diagrams \cite{GLM1}, we obtain 
the following expression for $\Omega_{i,k}(s,b)$:
\beq \label{FIMF}
\Omega^{i,k}_{\pom}\Lb Y; b\Rb\,\,\,= \,\,\, \int d^2 b'\,
\,\,\,\frac{ g_i\Lb\vec {b}'\Rb\,g_k\Lb\vec{b} - \vec{b}'\Rb
\,\Big(1/\gamma\, G_{\pom}\Lb T(Y)\Rb\Big)}
{1\,+\,\Lb G_{3\pom}/\gamma\Rb G_{\pom}\Big(T(Y)\Big)\,\left[
g_i\Lb\vec{b}'\Rb + g_k\Lb\vec{b} - \vec{b}'\Rb\right]}.
\eeq
$G_{3\pom}$ is the triple Pomeron vertex, 
and $\gamma^2 = \int \frac{d^2 k_t}{4 \pi^2} G^2_{3 \pom}$.
\subsubsection{Formulae for the double inclusive cross section}
Mueller diagrams for the different contributions to the 
double inclusive production are shown in \fig{CorGLM}. 
The diagram of \fig{CorGLM}-a is the same as we have discussed 
in section 3.1.  The main ingredient for this contribution is 
$\Gamma_i\Lb Y; b\Rb$, which is given by a slight modification of 
\eq{FUG}:
\beq \label{GA}
\Gamma_i \Lb Y - y, b\Rb\,\,=\,\,\frac{g_i\Lb b \Rb\frac{1}{\gamma}
G_\pom\Lb T\Lb Y - y\Rb\Rb}{1\,\,+
\,\,\Lb G_{3\pom}/\gamma\Rb\,g_i\Lb b \Rb G_\pom\Lb T\Lb Y - y\Rb\Rb}.
\eeq 
Introducing,
\beq
H_{i k}\Lb Y_1;Y_2; b\Rb\,\,\equiv\,\,\int d^2 b' 
\Gamma_i\Lb \vec{b} - \vec{b}',Y_1\Rb\,\Gamma_k\Lb \vec{b}',Y_2\Rb, 
\eeq 
we can rewrite the contribution of the diagram of \fig{CorGLM}-a in the form
\bea \label{IN2}
&&I_2\Lb y_1,y_2\Rb\,\,=\,\nonumber\\
&&\,a^2_{\pom}(\,\int d^2 b \left\{ \alpha^4
\exp\Lb \Omega_{11}\Lb \tilde{Y};b\Rb  - \Omega_{11}\Lb Y;b\Rb\Rb H_{11}\Lb 
\tilde{Y/}2 - y_1,\tilde{ Y}/2 + y_1;b\Rb\,H_{11}\Lb \tilde{Y}/2 - 
y_2, Y/2 + y_2;b\Rb \right.\nonumber\\
&&\left.
+2\alpha^2 \beta^2  \exp\Lb \Omega_{12}\Lb \tilde{Y};b\Rb  - 
\Omega_{12}\Lb Y;b\Rb\Rb H_{12}\Lb \tilde{Y/}2 - 
y_1,\tilde{ Y}/2 + y_1;b\Rb \,H_{12}\Lb \tilde{Y/}2 - 
y_1,\tilde{ Y}/2 + y_1;b\Rb \right.\nonumber\\
&& \left.+ \beta^4\,\exp\Lb \Omega_{22}\Lb \tilde{Y};b\Rb  - 
\Omega_{22}\Lb Y;b\Rb\Rb  H_{22}\Lb \tilde{Y/}2 - 
y_1,\tilde{ Y}/2 + y_1;b\Rb H_{22}\Lb \tilde{Y/}2 - 
y_1,\tilde{ Y}/2 + y_1;b\Rb \right\},
\eea
where, $ a_\pom$ is shown in \fig{CorCutP}. In \eq{IN2} we used the 
following notations: 
$Y\,=\,\ln\Lb s/s_0\Rb$ and $\tilde{Y} \,=
\,\,\ln\Lb \tilde{s}/s_0\Rb$. $y_1$ and $y_2$ are 
rapidities of the produced hadrons in the c.m.frame.
For the contribution of the  diagram of \fig{CorGLM}-b, 
we need to change $H_{i k}\Lb \tilde{Y/}2 - y_1,\tilde{ Y}/2 + y_1;b\Rb$ 
in \eq{IN2} to 
$J_{ik}\Lb y_1, y_2; b \Rb$ which is defined as
\bea \label{J}
J_{ik}\Lb y_1, y_2;  b \Rb\,\,&=&\,\,\int^{\tilde{Y}}_{\tilde{Y}/2 - y_1} 
d y'\,\int d^2 b' \,\Gamma_i\Lb \tilde{Y} - y'; \vec{b} - \vec{b}'\Rb\,
G_\pom\Lb T\Lb y' -\tilde{Y}/2 + y_1\Rb\Rb\,G_\pom\Lb T\Lb y' -\tilde{Y}/2 + 
y_2\Rb\Rb\nn\\ &\times &  \,\Gamma_k\Lb \tilde{ Y}/2 - y_1, b'\Rb
\,\Gamma_k\Lb \tilde{ Y}/2 - y_2, b'\Rb.
\eea
Therefore, this contribution takes the form
\bea \label{IN1}
&&I_1\Lb y_1,y_2\Rb\,\,=\,
\,a^2_{\pom}\,G_{3\pom}\,\int d^2 b \left\{ \alpha^4
\exp\Lb \Omega_{11}\Lb \tilde{Y};b\Rb  - 
\Omega_{11}\Lb Y;b\Rb\Rb\,J_{1 1}\Lb y_1, y_2; b \Rb  \right.\\
&&\left.
+2\alpha^2 \beta^2  \exp\Lb \Omega_{12}\Lb \tilde{Y};b\Rb  - 
\Omega_{12}\Lb Y;b\Rb\Rb\,J_{1 2}\Lb y_1, y_2; b \Rb + 
\beta^4\,\exp\Lb \Omega_{22}\Lb \tilde{Y};b\Rb  - 
\Omega_{22}\Lb Y;b\Rb\Rb  \,J_{2 2}\Lb y_1, y_2; b\Rb\, \right\}\nn.
\eea
Introducing,
\bea  \label{K}
K_{ik}\Lb y_1, y_2; b \Rb\,\,&=
&\,\,\int d^2 b' \,\int^{\tilde{Y}}_{ \tilde{Y/}2 - y_1} d y'\,\,
\Gamma_i\Lb \tilde{Y} - y'; \vec{b} - \vec{b}'\Rb\,
G_\pom\Lb T\Lb y' -\tilde{Y}/2 + y_1\Rb\Rb\,G_\pom\Lb T\Lb y' -
\tilde{Y/}2 + y_2\Rb\Rb\nn\\
&\times &  \,\int^{ \tilde{Y}/2 - y_2}_0 d y''\,\,\Gamma_k\Lb y''; 
\vec{b} - \vec{b}'\Rb\,
G_\pom\Lb T\Lb \tilde{Y}/2 - y_1 - y''\Rb\Rb\,G_\pom\Lb T\Lb  
\tilde{Y}/2 -  y_2 - y''\Rb\Rb.
\eea                                   
We can reduce the contribution of the diagram of \fig{CorGLM}-c to the 
form
\bea \label{IN3}
&&I_3\Lb y_1,y_2\Rb\,\,=\,
\,a^2_{\pom}\,G^2_{3\pom}\,\int d^2 b \left\{ \alpha^4
\exp\Lb \Omega_{11}\Lb \tilde{Y};b\Rb  - \Omega_{11}\Lb Y;b\Rb\Rb\,
K_{1 1}\Lb y_1, y_2; b \Rb  \right.\\
&&\left.
+2\alpha^2 \beta^2  \exp\Lb \Omega_{12}\Lb \tilde{Y};b\Rb  - 
\Omega_{12}\Lb Y;b\Rb\Rb\,K_{1 2}\Lb y_1, y_2; b \Rb + 
\beta^4\,\exp\Lb \Omega_{22}\Lb \tilde{Y};b\Rb  - 
\Omega_{22}\Lb Y;b\Rb\Rb  \,K_{2 2}\Lb y_1, y_2; b\Rb\, \right\}\nn.
\eea                                       

Collecting all contributions, the long range rapidity correlation 
function has the following form:

\bea
R(\eta_1,\eta_2)\,\,\,=\,\,\frac{\frac{h^2(\eta,Q)}
{\sigma_{in}(Y(\eta))}\left\{ I_1(y_1(\eta_1), y_1(\eta_2)) 
+I_2(y_1(\eta_1), y_1(\eta_2))+ I_3(y_1(\eta_1), y_1(\eta_2))\right\}} 
{\frac{1}{\sigma_{in}(Y)}\frac{d \sigma}{d y_1}\,
\frac{1}{\sigma_{in}(Y)}\frac{d \sigma}{d y_1}}\,\,-\,\,1.
\eea                                                                         
 Expressions for  the single inclusive cross section 
$\frac{1}{\sigma_{in}(Y)}\frac{d \sigma}{d y_i}$ 
as well as the Jacobian $h$ and the definition of the pseudo-rapidity 
$\eta$ can be found in Ref.\cite{GLMIN}.
\subsubsection{Correlations in the GLM model}
\begin{figure}[h]
\begin{center}
\includegraphics[width=12cm]{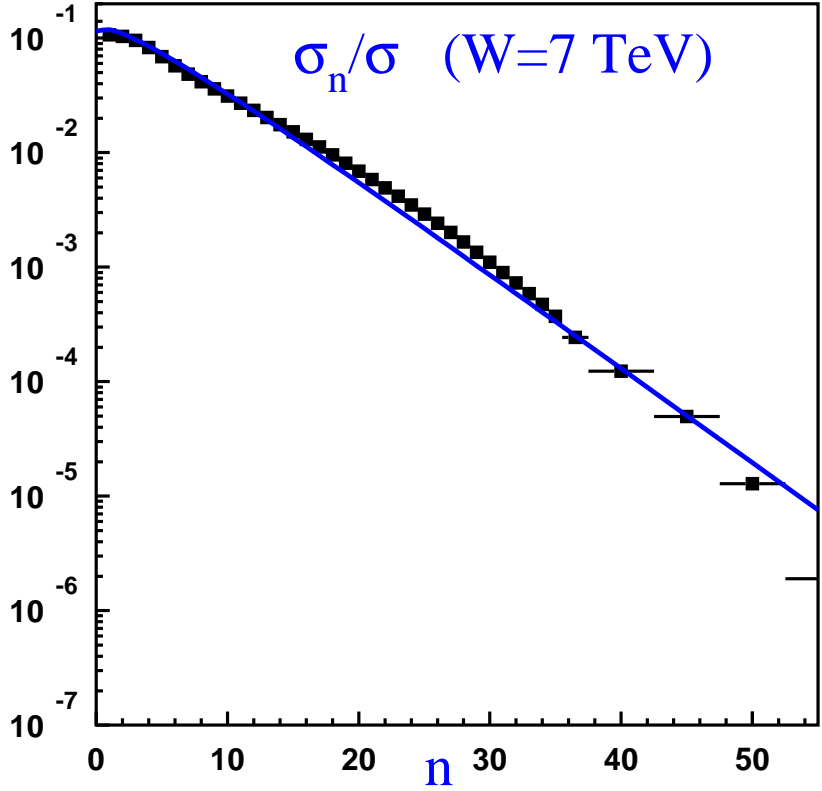}
\end{center}
\caption{Multiplicity distribution measured by CMS collaboration 
\cite{CMSMULT} and \protect\eq{NBD} with our parameters.}
\label{sn}
\end{figure}

\begin{figure}[h]
\begin{tabular}{c c}
\includegraphics[width=8.5cm]{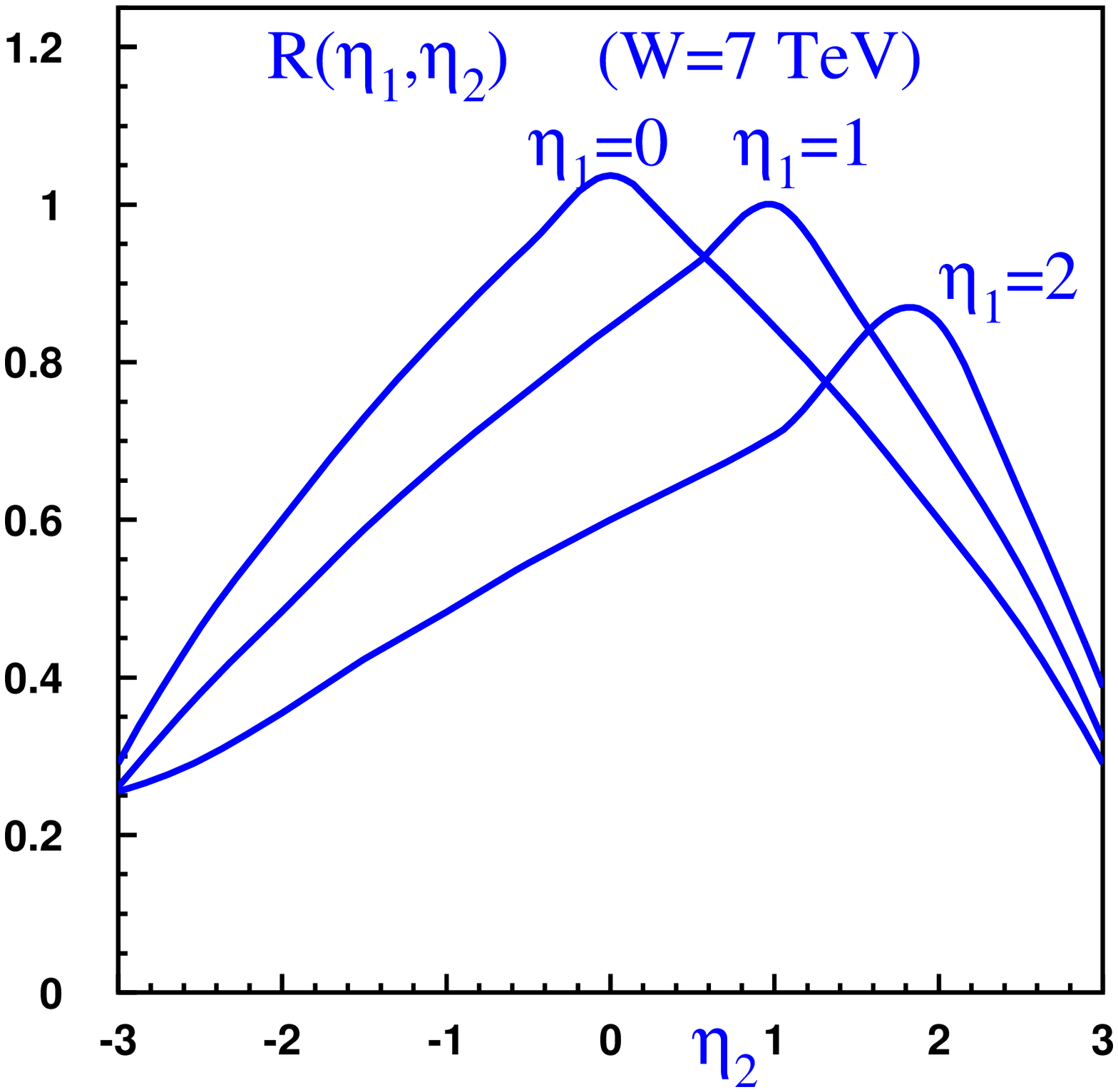}
&\includegraphics[width=8.5cm]{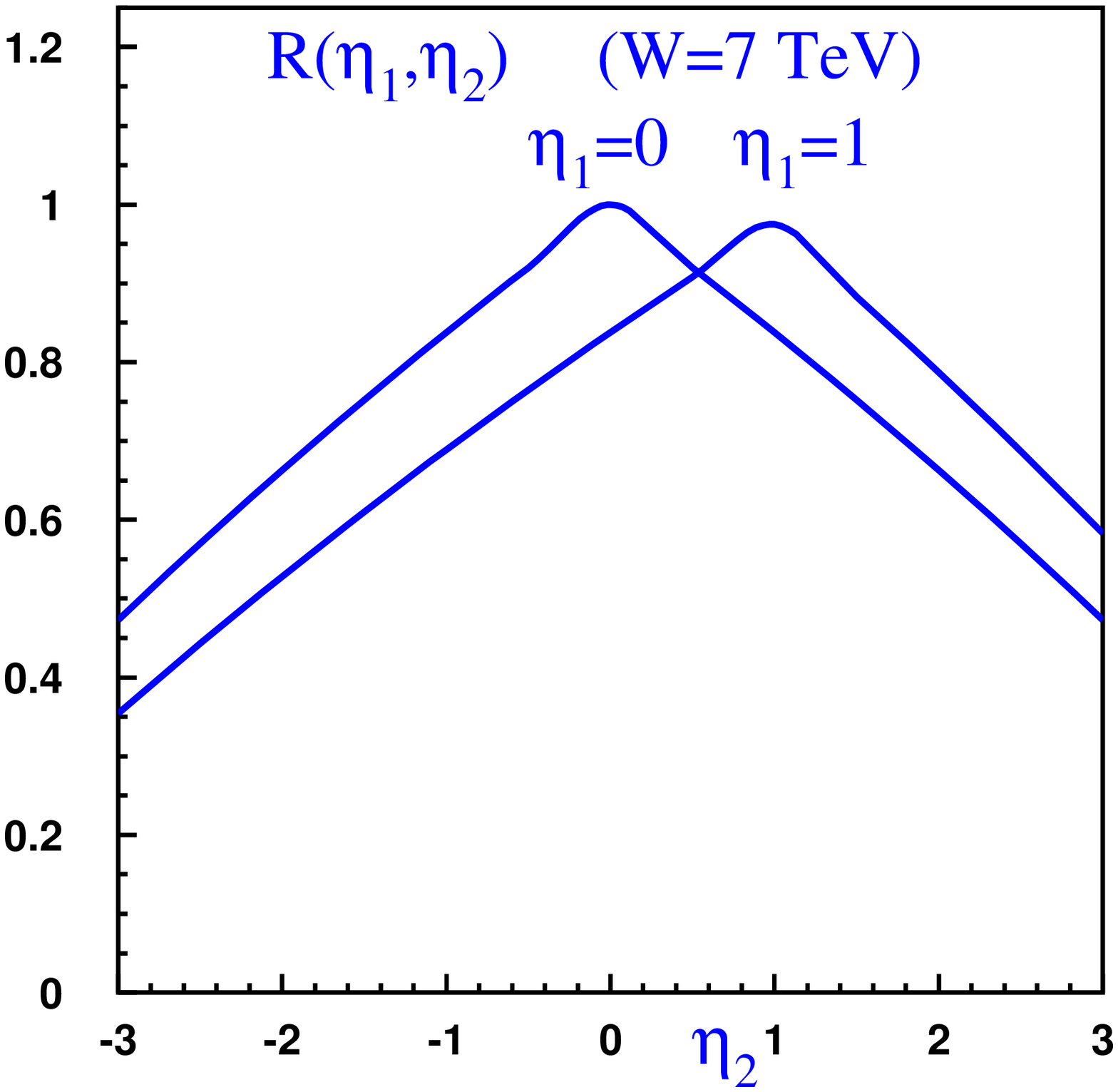}\\
\fig{cory}-a & \fig{cory}-b\\
\end{tabular}
\caption{Our prediction for $R\Lb \eta_1, \eta_2\Rb$ versus $\eta_2$ 
at different values of $\eta_1$ at $W = 7 \,TeV$ (\protect\fig{cory}-a) and the 
estimates of the simple model (see \protect\eq{SIMO} with $ \Delta_\pom = 0.08$) 
for the correlation function (\protect\fig{cory}-b). }
\label{cory}
\end{figure}

Using the formulae of the previous section we calculate the correlations in the
 GLM model. It turns out that  $R\Lb 0,0\Rb$ is a constant in the energy range 
W = 0.9 to 7  TeV, and it is equal to $R\Lb 0,0 \Rb \approx 2$ (see Table 
2
).
\begin{table}
\begin{center}
\begin{tabular}{|c|c|c|c|c|}
\hline \hline
 W(TeV) & 0.9 & 1.8 & 2.36 &7\\
 \hline
$R(y_1 = 0,y_2 = 0)$& 1.0 & 1.12 & 1.026& 1.034\\
\hline\hline
\end{tabular}
\end{center}
\caption{$R\Lb y_1=0, y_2=0\Rb$ versus energy.}
\end{table}
This result is in a good agreement with the CMS data on 
multiplicity distribution \cite{CMSMULT}. Indeed, experimentally, 
$C_2 \,=\,\langle n^2 \rangle/\langle n \rangle^2$ 
was measured for the rapidity window $|\eta|\, < \,0.5$ 
in the energy range W = 0.9 to  7 TeV 
(see Fig.6 in Ref.\cite{CMSMULT}) and $C_2 \approx 2$. 
For this small range of rapidity, 
we can consider that $C_2 = R\Lb 0, 0\Rb + 1$. 
It is worthwhile mentioning that using our calculation of $R\Lb0,0\Rb$, 
we can calculate the parameters of the negative binomial distribution
\beq \label{NBD}
\frac{\sigma_n}{\sigma_{in}}\,\,=\,\,\Lb \frac{r}{r + 
\langle n \rangle}\Rb^r 
\frac{\Gamma\Lb n + r\Rb}{n!\, \Gamma\Lb r \Rb}\Lb 
\frac{\langle n \rangle}{r\,+\,\langle n \rangle}\Rb^n.
\eeq
In our model, given $|\eta| \leq 0.5$, 
$\langle n \rangle \,=\,$5.8 (see Ref.\cite{GLMIN}) and $r = 1.25$.
Using this distribution we calculate 
$C_q\,=\,\langle n^q\rangle/\langle n \rangle^q$. 
They equal $ C_3 $ = 5.65, $C_4\,=\,21.18$ and $C_4$=98.2. They are in 
good agreement  with the experimental data of Ref.\cite{CMSMULT} 
except $C_4$ which experimentally is about 70. In 
\fig{sn} we compare \eq{NBD} with the CMS experimental data at $W=7\,TeV$.
In \fig{cory} we plot the correlation function $R\Lb \eta_1,\eta_2\Rb$ 
as a function of $\eta_2$. One can see that this function falls
steeply  at large $\eta_2$. At first sight, such form of $\eta_2$ 
dependence looks strange since all diagrams of \fig{CorGLM} 
generate long range rapidity correlations. 
It turns out that the main contribution comes from the enhanced diagram of 
\fig{CorGLM}-c. The eikonal-type diagram of \fig{CorGLM}-a 
leads to long range rapidity correlations which do not depend on the values 
of $\eta_1$ and $\eta_2$. The diagram of \fig{CorGLM}-b gives 
a negligible contribution.
Let us consider \fig{CorGLM}-c in a simple model replacing $\Gamma(y) $ by 
the exchange of the Pomeron, and considering all Pomeron exchanges
as the exchange of a `bare' Pomeron. 
In this model the diagram of \fig{CorGLM} has the form:
\bea \label{SIMO}
&&g^2_p a^2_\pom  G^2_{3 \pom}\int^Y_{y_1} d y'' 
\int^{y_2}_0 d y"\,G_\pom\Lb Y - y'\Rb \,G_\pom\Lb y' - y_1\Rb G_\pom\Lb 
y' - y_1\Rb G_\pom\Lb y' - y_2\Rb \nn\\
&&\times\,\,~~~
G_\pom\Lb y_2 - y" \Rb G_\pom\Lb y_2 - y" \Rb G_\pom\Lb  y" \Rb=\nn\\
&& g^2_p  a^2_\pom \frac{G^2_{3 \pom} }{\Delta^2_\pom} 
e^{2 \Delta_\pom Y}\Big(1 - e^{\Delta_\pom (y_1 - Y)} - e^{\Delta_\pom 
(y_1 - Y)} -  e^{\Delta_\pom (- y_2 )}  +  
e^{\Delta_\pom (y_1 - y_2  - Y)} \Big),
\eea
where, we used $G_\pom\Lb Y\Rb = \exp\Lb \Delta_\pom Y\Rb$. Recalling 
that the single inclusive cross section 
$d \sigma/d y = g^2_p a^2_p \exp\Lb \Delta_\pom Y\Rb$, 
 in this simple model, 
the correlation function of \eq{I1} is equal to
\beq
\label{SIMO1}
R\Lb y_1, y_2 \Rb\,\,=\,\,\sigma_{in}\frac{G^2_{3 \pom}}{\Delta^2_\pom} 
\Big(1 - e^{ \Delta_\pom (y_1 - Y)} - e^{\Delta_\pom (y_1 - Y)} - 
e^{ \Delta_\pom  (- y_2  )}  +  e^{ \Delta_\pom (y_1 - y_2  - Y )} 
\Big)\,\,-\,\,1.
\eeq
In \fig{cory}-b the correlation function is plotted with 
$\sigma_{in} G_{3\pom}/\Delta^2_\pom = 2$ 
and $\Delta_\pom = 0.08$ which correspond to the effective behaviour of 
the dressed Pomeron in our model at high energies ($ W = 1.8 - 7 TeV$). 
One can see that simple formula of \eq{SIMO1} reproduces 
the short-range correlation type  behaviour  of \fig{cory}-a.
\section{Conclusions}
In this paper we taken the next step, following the single inclusive cross 
section\cite{GLMIN}, in the description of the multi particle production 
processes in the framework of our soft interaction model.  
The main ingredients of our model are the large Pomeron intercept 
($\Delta_\pom = 0.23$ ) and  $\alpha'_\pom = 0$. 
 The model gives 
a practical realization of the BFKL Pomeron Calculus 
in zero transverse dimensions. The model reproduces quite well all 
classical soft scattering data: total, elastic and diffractive cross 
sections and the energy dependence of the elastic slope in wide range of 
energy W = 20 GeV to  7 TeV. 
The attraction of the Pomeron approach reveals itself in the possibility 
to discuss not only the forward scattering data but, also,  
to make predictions relating to multiparticle production processes
using the AGK cutting rules \cite{AGK}.

In this paper we have developed a procedure for  calculating  the 
correlation function in the MPSI approximation utilizing 
the BFKL Pomeron Calculus in zero transverse dimensions. 
The  theoretical formulae obtained allow us to calculate the 
rapidity correlation function in our model for soft interactions. We 
compare our prediction with the multiplicity distribution at 
$W\, =\, 7\,$ TeV measured by CMS collaboration \cite{CMSMULT}, 
which we describe quite well. 
In \fig{cory} we present our prediction for the rapidity dependence of 
the correlation function.

We believe that our approach opens the way to discuss the structure of the 
bias events without building Monte Carlo codes.
At the moment we demonstrate 
that our model describes all standard soft data on forward scattering, 
inclusive cross sections and multiplicity distribution. 
We also predict the rapidity correlation function.

We thank all participants of ``Low x'2013 WS'' 
for fruifful discussions on the subject.
This research of E.L. was supported by the Fondecyt (Chile) grant 1100648.

\end{document}